\documentclass[aps,prb,reprint,amsmath,amssymb,groupedaddress]{revtex4-2}

\usepackage{graphicx}
\usepackage{dcolumn}
\usepackage{bm}
\usepackage{color}
\usepackage{siunitx}
\usepackage{todonotes}
\usepackage{threeparttable}
\usepackage{gensymb}
\usepackage{ulem}

\definecolor{imcolor}{rgb}{0.5,0.,0.5}				

\definecolor{ohcolor}{rgb}{0.,0.5,0.5}

\definecolor{hpcolor}{rgb}{0.5,0.5,0}

\begin{document}

\title{DFT+DMFT study of the magnetic susceptibility and the correlated electronic structure in transition-metal intercalated NbS$_2$}

\author{Hyowon Park$^{1,2}$ and Ivar Martin$^{1}$}

\affiliation{$^1$Materials Science Division, Argonne National Laboratory, Argonne, IL, 60439, USA,\\
$^2$Department of Physics, University of Illinois at Chicago, Chicago, IL 60607, USA }

\date{\today}

\begin{abstract}
The Co-intercalated NbS$_2$ (Co$_{1/3}$NbS$_2$) compound exhibits  large anomalous  Hall conductance, likely due to the non-coplanar magnetic ordering  of Co spins. In this work, we study the relation between this novel magnetism and the correlated electronic structure of Co$_{1/3}$NbS$_2$ by adopting dynamical mean field theory (DMFT) to treat the correlation effect of Co $d$ orbitals. 
We find that the hole doping  of Co$_{1/3}$NbS$_2$ can tune the size of the Nb hole pocket at the DMFT Fermi surface,   producing features consistent with those observed in angle resolved photoemission spectra [Phys. Rev. B \textbf{105}, L121102 (2022)].
We also compute the momentum-resolved spin susceptibility, and correlate it with   
the Fermi surface shape. 
We find that the magnetic ordering wavevector of Co$_{1/3}$NbS$_2$ obtained from the peak in spin susceptibility agrees  with the one observed experimentally by neutron scattering; it is compatible with commensurate non-coplanar $3q$ spin structure.
We also discuss how results change if some other than Co  transition metal intercalations are used.
\end{abstract}

\maketitle

\section{Introduction}

Understanding the relation between a novel electronic transport and the correlated electronic structure of complex materials has been a grand challenge in condensed matter physics. For instance,
cobalt-intercalated NbS$_2$ (Co$_{1/3}$NbS$_2$) shows a very large anomalous  Hall effect~\cite{ghimire_large_2018,PhysRevResearch.2.023051}; however, the origin of this  phenomenon has remained a subject of debate.
While the `standard' anomalous Hall effect originates stems from finite  uniform magnetization in ferromagnets \cite{RevModPhys.82.1539},  the antiferromagnetic (AFM) ground state of Co$_{1/3}$NbS$_2$ implies a different, more  exotic origin.  
One promising scenario is that topologically non-trivial magnetism such as the non-coplanar spin state of Co $d$ orbitals can generate a strong fictitious magnetic field, thus inducing a ``topological"  Hall effect \cite{martin_scalar_2008}. Equivalently, in the band language, the Co spin moments couple to the itinerant Nb bands and produce the topologically non-trivial band structure resulting in a large Berry curvature leading to  anomalously large Hall currents. 

Our previous first-principle calculations based on density functional theory (DFT) support this scenario: the energy of the non-coplanar $3q$ magnetic structure  is the lowest, compared to other $1q$ or $2q$ states~\cite{PhysRevMaterials.6.024201}. Moreover, the Berry phase calculation based on the magnetic band structure supports a large anomalous Hall conductivity (AHC), comparable to $e^2/h$ per crystalline layer~\cite{PhysRevMaterials.6.024201}.
Unfortunately, the DFT band structure of Co$_{1/3}$NbS$_2$  fails to capture the angle-resolved photoemission spectra (ARPES), which requires going beyond the rigid-band shift picture due to the Co intercalation~\cite{PhysRevB.105.L121107,PhysRevB.105.155114}.
One important feature observed in  ARPES measurements in Co$_{1/3}$NbS$_2$ is the appearance of
the broad electron pocket around the high symmetry $K$ point, which is not  captured by DFT~\cite{PhysRevB.105.L121107,PhysRevB.105.155114,PhysRevB.105.L121102}. 
Moreover, the effective electron mass at the electron pocket of Co$_{1/3}$NbS$_2$ is twice larger than that of NbS$_2$~\cite{PhysRevB.105.155114}.
These suggest that for a more accurate picture, one needs to treat the strong correlation effect of Co $d$ orbitals beyond DFT.

Until recently, experimental support for  exotic magnetic states in Co$_{1/3}$NbS$_2$ had been lacking.
In fact, early neutron scattering measurement on Co$_{1/3}$NbS$_2$ argued that the scattering peak data fits well to the standard commensurate ($1q$) AFM structure, but with multiple magnetic domains~\cite{parkin_magnetic_1983}. (One should note, that it is quite difficult to distinguish between multi-domain $1q$  AFM state and a mono-domain $3q$ state in neutron scattering.)
Recently, however, polarized neutron scattering measurements on the related material  Co$_{1/3}$TaS$_2$, which has the same structure but with Nb replaced by Ta, convincingly demonstrated the presence of non-coplanar  Co magnetism. Moreover, they demonstrated connection between the appearance of this magnetic state and large spontaneous topological Hall effect ~\cite{NaturePhysics_Takagi}.

What is the physical origin of noncoplanar magnetism? In pure-spin models it typically requires having multi-spin interactions (either four or six spin). When interactions are mediated by itinerant electrons, such higher order terms are generated naturally \cite{batista2016frustration}. From the weak-coupling perspective, having multiple $q$ orders present simultaneously allows to gap out larger total sections of Fermi.
The key ingredients for noncoplanar order are (1) the large susceptibility with respect to simple collinear single-$q$ order (e.g. if $q$ connect nearly flat opposite sides of the Fermi surface -- nesting effect), and (2) at least approximate commensuration of the magnetic order and the crystal lattice. The latter criterion  allows for several $q$ orderings  to coexist with each other without suppressing the local amplitude of the magnetic order.

In this work, we focus on the first element, the analysis of magnetic susceptibility. We go beyond the standard DFT by adopting dynamical mean field theory (DMFT), which allows to treat the strong correlation effect on the transition metal ($M$) $d$ orbitals in $M_{1/3}$NbS$_2$. As the first step, we match the main features of DMFT Fermi surface calculations to be consistent with the ARPES data. 
We then calculate the momentum-dependent magnetic susceptibility $\chi$ from first-principles and investigate the momentum $q$ vector showing the leading instability.
We note that DFT alone is not sufficient to study the correlated electronic structure of Co$_{1/3}$NbS$_2$ as it fails to capture the essential features of the APRES measurement.
Also, compared to our previous DFT study on the Co$_{1/3}$NbS$_2$ that showed  $3q$ non-coplanar structure to be the lowest in energy compared to possible $1q$ or $2q$ states, now we are allowing for the possibility of instability at a wavevectors incommensurate with the lattice.


\section{Methods}
In this section, we explain computational methods used in the band structure and the magnetic susceptibility calculations of $M$Nb$_3$S$_6$ ($M$= Co, Fe, and Ni). We also provide parameters used in the calculations.

\subsection{DMFT calculation}

To study the band structure and the Fermi surface of $M$Nb$_3$S$_6$ ($M$= Co, Fe, and Ni), we adopt DFT+DMFT treating the strong correlation effect of $M$ ions.
The procedure of the DMFT calculation is as follows.
First, we obtain the non-spin-polarized (nsp) band structure from the experimental $M$Nb$_3$S$_6$ crystal structures.
We adopted the Vienna Ab-initio Simulation Package (VASP)~\cite{vasp1,vasp2} code to compute the nsp band structure using a $14\times14\times4$ $k-$mesh along with the energy cutoff of 400eV for the plane-wave basis.  
We used the Perdew-Burke-Ernzerhof (PBE) functional for the exchange and correlation energy of DFT. 
Using the nsp band structure, we construct the following tight-binding Hamiltonian by adopting the maximally localized Wannier function~\cite{marzari1997maximally} as the basis,
\begin{eqnarray}
\label{eq:Ham}
\hat{H}&=&\sum_{\alpha\beta,ij\sigma}t_{\alpha\beta, ij} \hat{c}_{i\alpha\sigma}^{\dagger}\hat{c}_{j\beta\sigma}+\sum_{\alpha\beta,i}\sum_{\sigma\sigma'}U^{\sigma\sigma'}_{\alpha\beta}\hat{n}_{i\alpha\sigma}\hat{n}_{i\beta\sigma'} \\
&&+\sum_{ij\sigma}t_{ij} \hat{d}_{i\sigma}^{\dagger}\hat{d}_{j\sigma}+\sum_{\alpha,ij\sigma} (t_{\alpha,ij}\hat{c}_{i\alpha\sigma}^{\dagger}\hat{d}_{j\sigma}+t^*_{\alpha,ij}\hat{d}_{j\sigma}^{\dagger}\hat{c}_{i\alpha\sigma}),
\nonumber 
\end{eqnarray}
where $t$ is the inter-orbital and inter-site hopping matrix elements including the whole manifold of the Co $d$ orbitals ($\hat{c}^{\dagger}, \hat{c}$) and the Nb $d_{z^2}$ orbital ($\hat{d}^{\dagger}, \hat{d}$).
$\hat{n}_{i\alpha\sigma} (=\hat{c}^{\dagger}_{i\alpha\sigma}\hat{c}_{i\alpha\sigma})$ is the density operator for the orbital $\alpha$ and the spin $\sigma$ at the site $i$.
$U^{\sigma\sigma'}_{\alpha\beta}$ is the local Coulomb interaction matrix for the on-site Co $d$ orbitals and approximated as the density-density interaction type.

Using the Hamiltonian in Eq.\:\ref{eq:Ham}, we solve the DMFT self-consistent equations~\cite{singh2021dmftwdft} using the continuous-time quantum Monte Carlo (CTQMC) method~\cite{ctqmc1} as the impurity solver, then obtain the the local self-energy $\Sigma(i\nu_n)$ for the Co $d$ orbitals. 
Here, we parameterize the $U_{\alpha\beta}$ matrix elements by the Slater integrals using
the local Hubbard interaction $U$=5eV and the Hund's coupling $J$=0.7eV. The temperature $T$ is set to be 116K.
In DMFT, we use a fine $k-$mesh of 30$\times$30$\times$10.
For all compounds, the fixed double counting potential scheme is adopted using the following formula to subtract the double-counting correction from the DFT potential
\begin{equation}
\label{eq:double_counting_nominal}
    V_{DC}=U(n_d^0-\frac{1}{2})-\frac{J}{2}(n_d^0-1)
\end{equation}
where $n_d^0$ is the nominal occupancy of the transition metal $d$ orbitals, i.e. $n_d^0=7.0$ for Co$^{2+}$, $n_d^0=6.0$ for Fe$^{2+}$, and $n_d^0=8.0$ for Ni$^{2+}$.

\subsection{Spin susceptibility $\chi^{sp}(\mathbf{q})$ calculation}

A general form of the spin susceptibility $\chi^{sp}$ can be given by the retarded two-particle Green's function of two spin operators:
\begin{equation}
\chi_{mn}^{sp}(\mathbf{r}'-\mathbf{r},t'-t) =
\Theta(t'-t)\langle [\hat{S}^m(\mathbf{r}',t'), \:
\hat{S}^n(\mathbf{r},t) ]_-\rangle,
\end{equation}
where ${\hat{S}^n(\mathbf{r},t)}$ is the $n$-th component of spin density
and $\Theta(t'-t)$ is the step function that imposes causality.
The spin operator can be expanded using the localized orbital basis set and the fermionic creation/annihilation operators:
\begin{equation}
\hat{S}^n(\mathbf{r},t) = \frac{1}{N}
\sum_{\sigma\sigma'}\hat{s}^n_{\sigma\sigma'} \sum_{n n'}   \phi^{*}_n(\mathbf{r})\phi_{n'}(\mathbf{r}) \:\hat{c}^{\dagger}_{n\sigma}(t) \hat{c}_{n'\sigma'}(t),
\end{equation}
where 
$\hat{s}^i=g\mu_B\hat{\sigma}^i/2$ with $\hat{\sigma}^i$ being the $i$-th Pauli matrix and $\phi_{n}(\mathbf{r})$ is the basis function with the index $n \:(=\{\mathbf{k},\alpha,\tau_{\alpha}\})$ for the orbital character $\alpha$ located at the position $\tau_{\alpha}$ with the momentum $\mathbf{k}$. One can note that the spin operator can be diagonal for both spin and orbital basis sets if the spin arrangement is collinear and the spin-orbit coupling is neglected. 
Here, we consider the paramagnetic spin symmetry without the spin-orbit coupling. 


For the longitudinal and paramagnetic spin symmetries ($m=n=z$),
the spin susceptibility $\chi_{zz}^{sp}$ 
can be obtained from 
\begin{eqnarray}
\label{eq:chisp}
\chi_{zz}^{sp}(\mathbf{r}'-\mathbf{r},t'-t) &=&  \frac{\Theta(t'-t)}{N^2}\cdot\sum_{\sigma\sigma'}\sum_{\bar{\sigma}\bar{\sigma}'}\hat{s}^z_{\sigma\sigma'}\hat{s}^z_{\bar{\sigma}\bar{\sigma}'} \\
&& \sum_{nn'}\sum_{\bar{n}\bar{n}'} \phi^*_{n}(\mathbf{r'})\phi_{n'}(\mathbf{r'})
\phi^*_{\bar{n}}(\mathbf{r})\phi_{\bar{n}'}(\mathbf{r})\nonumber \\ 
 && \cdot \langle [\hat{c}^{\dagger}_{n\sigma}(t') \hat{c}_{n'\sigma'}(t'), \:
\hat{c}^{\dagger}_{\bar{n}\bar{\sigma}}(t) \hat{c}_{\bar{n}'\bar{\sigma}'}(t) ]_-\rangle. \nonumber
\end{eqnarray}
Here, the paramagnetic symmetry imposes that the two-particle response function should be invariant upon the spin flip, i.e. $\sigma\rightarrow -\sigma$.



\subsection{Form factor $F$ calculation}

Using the paramagnetic symmetry, the momentum and frequency dependent susceptibility, $\chi_{zz}^{sp}(\mathbf{q},\omega)$ can be simplified from the Fourier transform of Eq.\:\ref{eq:chisp}:
\begin{equation}
\chi_{zz}^{sp}(\mathbf{q},\omega) = \frac{(g\mu_B)^2}{2N^2} \sum_{nn'}\sum_{\bar{n}\bar{n}'} F_{nn'\bar{n}\bar{n}'}(\mathbf{q})\cdot\chi_{nn'\bar{n}\bar{n}'}\left(\omega\right),
\label{eq:chiqw}
\end{equation}
where $F(\mathbf{q})$ is the atomic form factor describing the modulation of the charge density and
$\chi(\omega)$ is the orbital-dependent two-particle response function where the spin dependence is simplified due to the paramagnetic symmetry.
In the susceptibility calculation based on DFT, the matrix element for $F(\mathbf{q})$ can be typically computed using the Kohn-Sham wavefunction, i.e. $\psi_{n_{\alpha}}^{\mathbf{k}}(\mathbf{r})$ (the orbital index $\{\alpha,\tau_{\alpha}\}$ changes to the band index $n_{\alpha}$), and it can be expanded continuum basis set, such as plane waves~\cite{PhysRevB.90.115143}. 
Here, we adopt the maximally localized Wannier function for the form factor $F(\mathbf{q})$ calculation, which is the same basis function for DMFT calculations:
\begin{eqnarray}
\phi_{n}(\mathbf{r})=\phi^{\mathbf{k}}_{n_{\alpha}}(\mathbf{r})=\frac{1}{\sqrt{N}}\sum_{\mathbf{R}} e^{-i\mathbf{k}\cdot\mathbf{R}}\phi^{\tau_{\alpha}}_{\alpha}(\mathbf{r}-\mathbf{R}),
\end{eqnarray}
where $\phi^{\mathbf{k}}_{n_{\alpha}}(\mathbf{r})$ is the Wannier function with the index $n_{\alpha}\:(=\{\alpha,\tau_{\alpha}\})$ defined in the $\mathbf{k}-$space for the primitive unit cell.
The index $n_{\alpha}$ runs over both the orbital character $\alpha$ and the internal atomic position $\tau_{\alpha}$.

If the complete and orthonormalized basis set is used for the $F(\mathbf{q})$ calculation, one can obtain the product of delta functions imposing the momentum and orbital conservation:
\begin{eqnarray}
F_{nn'\bar{n}\bar{n}'}(\mathbf{q})&=& \delta_{\mathbf{k}+\mathbf{q},\mathbf{k'}} \cdot 
\delta_{\bar{\mathbf{k}}'+\mathbf{q},\bar{\mathbf{k}}} \cdot \delta_{n_{\alpha},n_{\alpha'}} \cdot \delta_{n_{\bar{\alpha}},n_{\bar{\alpha}'}}.
\label{eq:delta_form}
\end{eqnarray}
However, in general, $F(\mathbf{q})$ needs to be modified if one uses the incomplete basis set.
In the case of Co$_{1/3}$NbS$_2$, the magnetic moments  primarily reside on Co $d$ orbitals, which are the subset of the complete band structure. 
As a result, the $F(\mathbf{q})$ expression for only Co $d$ orbitals can be modified as follows:
\begin{eqnarray}
F_{nn'\bar{n}\bar{n}'}(\mathbf{q})&=&\frac{1}{N^2_{\tau}}\biggl(\:\sum_{\mathbf{\tilde{k}},\mathbf{\tilde{k}}'} \sum_{\vec{\tau}_\alpha,\vec{\tau}_{\alpha'}} e^{i(\tilde{\mathbf{k}}'\cdot\vec{\tau}_{\alpha'}-\tilde{\mathbf{k}}\cdot\vec{\tau}_{\alpha})}
 \nonumber\\
&&\cdot\delta_{\alpha \alpha'}\cdot\delta_{\{\mathbf{k}\},\tilde{\mathbf{k}}}
\cdot \delta_{\{\mathbf{k}'\},\tilde{\mathbf{k}}'}
\cdot \delta_{\tilde{\mathbf{k}}+\{\mathbf{q}\}, \tilde{\mathbf{k}}'}\biggr) \nonumber\\
&& \cdot\biggl(\:\sum_{\mathbf{\tilde{k}},\mathbf{\tilde{k}}'} \sum_{\vec{\tau}_{\bar{\alpha}},\vec{\tau}_{\bar{\alpha}'}} e^{i(\tilde{\mathbf{k}}'\cdot\vec{\tau}_{\bar{\alpha}'}-\tilde{\mathbf{k}}\cdot\vec{\tau}_{\bar{\alpha}})} \nonumber\\
&&\cdot\delta_{\bar{\alpha} \bar{\alpha}'}\cdot\delta_{\{\bar{\mathbf{k}}\},\tilde{\mathbf{k}}}
\cdot \delta_{\{\bar{\mathbf{k}}'\},\tilde{\mathbf{k}}'}
\cdot \delta_{\tilde{\mathbf{k}}'+\{\mathbf{q}\}, \tilde{\mathbf{k}}}\biggr),
\label{eq:form_final}
\end{eqnarray}
where the momentum $\tilde{\mathbf{k}}$ is defined in an extended Brillouin zone (BZ) obtained for a single Co ion in triangular lattice, $\{\mathbf{k}\}$ represents the set of the momentum $\mathbf{k}$ vectors shifted by the reciprocal vectors, $\vec{\tau}_\alpha$ is the atomic position of correlated atoms, and $N_{\tau}$ is the number of correlated atoms in the primitive unit cell. More detailed derivation of Eq.\:\ref{eq:form_final} is given in the Appendix.

\subsection{The Bethe-Salpeter equation}

The orbital-dependent susceptibility $\chi_{nn'\bar{n}\bar{n}'}(\omega)$ in Eq.\:\ref{eq:chiqw} can be computed using the following Bethe-Salpeter equation:
\begin{equation}
\chi_{nn'\bar{n}\bar{n}'}(\omega) = \chi_{n_\alpha n_{\bar{\alpha}}}^0(\mathbf{k},\mathbf{k'},\omega) + \chi^0*\Gamma^{irr}*\chi,
\label{eq:BSE}
\end{equation}
where $\chi^0$ is the polarizability obtained from the interacting Green's function and $\Gamma^{irr}$ is the irreducible vertex function.
One should note that the orbital indices for $\chi^0_{n_\alpha n_{\bar{\alpha}}}$ run over all Co $d$ and Nb $d_{z^2}$ orbitals in a unit cell while those on the $F(\mathbf{q})$ in Eq.\:\ref{eq:form_final} account only the correlated Co $d$ orbitals. 
In general, $\Gamma^{irr}$ is a complex function depending on momentum, frequency, spin, site, and orbital degrees of freedom.
While the effect of $\Gamma^{irr}$ is crucial to compare both momentum and frequency dependence of the susceptibility to the experimental neutron scattering data~\cite{PhysRevLett.107.137007,doi:10.1126/science.aan0593}, we adopt the static interaction type based on the random phase approximation (RPA), namely assuming that the interaction matrix to be independent on momentum and frequency.
Here, we further approximate that it is independent of orbitals to account for the average interaction effect and has the spin rotation symmetry to consider only interactions in the spin channel. 
As a result, we consider both the on-site interaction $\bar{U}$ within Co ions and the inter-site interaction $\bar{V}$ between Co and Nb ions, then study the effects of $\bar{U}$ and $\bar{V}$ on the susceptibility calculations.
For the on-site $\bar{U}$ value, we used the $\bar{U}$=1.5eV, which is smaller than the DMFT $U$ value. 
This is because the static two-particle interaction is further renormalized within the RPA diagrams while the diagrams within DMFT take the orbital and dynamical fluctuations into account explicitly.
We also used different inter-site $\bar{V}$ values ($\bar{V}$=0, 0.2, and 0.3 eV) to explore the effect of $\bar{V}$ on the susceptibility.
Our results show that the inter-site $\bar{V}$ can enhance the momentum dependence of the susceptibility significantly.

The polarizability $\chi^0$ can be given by the product of two Green's functions using the Wick's theorem. 
Our $\chi^0$ is different from the bare susceptibility since it is obtained from the interacting Green's function.
In DMFT, the Green's function is dressed with the local self-energy and the Matsubara frequency sum over $i\nu$ can be evaluated by performing the contour integral to obtain the polarizability at $\omega=0^+$:
\begin{eqnarray}
    \chi^{0\:\mathbf{k},\mathbf{k'}}_{n_\alpha n_{\bar{\alpha}}}&=&-T\sum_{i\nu}G_{n_\alpha n_{\bar{\alpha}}}(\textbf{k},i\nu)\cdot G_{n_{\bar{\alpha}} n_{\alpha}}(\textbf{k}',i\nu+i0^+) \\
 & = & -\frac{T}{2\pi i}\oint dz \:\: G_{n_\alpha n_{\bar{\alpha}}}(\textbf{k},z)\cdot G_{n_{\bar{\alpha}} n_{\alpha}}(\textbf{k}',z+i0^+)\nonumber \\
 & = & \frac{1}{\pi}\int d\nu\:[ImG_{n_\alpha n_{\bar{\alpha}}}(\textbf{k},\nu)\cdot G_{n_{\bar{\alpha}} n_{\alpha}}(\textbf{k}',\nu+i0^+)\nonumber \\ 
 && +G_{n_\alpha n_{\bar{\alpha}}}(\textbf{k},\nu-i0^+)\cdot ImG_{n_{\bar{\alpha}} n_{\alpha}}(\textbf{k}',\nu)]f(\nu)
 ,\nonumber
  \label{eq:chi0}
\end{eqnarray}
where $f(\nu)$ is the Fermi function.
The real part of the susceptibility at $\omega=0^+$ is given by
\begin{eqnarray}
Re\chi^{0\:\mathbf{k},\mathbf{k'}}_{n_\alpha n_{\bar{\alpha}}} & = & \frac{1}{\pi}\int d\nu f(\nu)[ImG_{n_\alpha n_{\bar{\alpha}}}(\textbf{k},\nu)\cdot ReG_{n_{\bar{\alpha}} n_{\alpha}}(\textbf{k}',\nu)\nonumber \\ 
 && +ReG_{n_\alpha n_{\bar{\alpha}}}(\textbf{k},\nu)\cdot ImG_{n_{\bar{\alpha}} n_{\alpha}}(\textbf{k}',\nu)]
 ,
 \label{eq:Rechi0}
\end{eqnarray}
where both $ReG$ and $ImG$ are the real and imaginary parts of interacting Green's functions defined on the fine $\mathbf{k}$ mesh and the real frequency which are obtained from the analytic continuation of the DMFT self-energy using the maximum entropy method.
For the $\chi^0$ susceptibility calculation in Eq.\:\ref{eq:chi0}, we performed the summation over the dense $\mathbf{k}-$grid using 60$\times$60$\times$10 $\mathbf{k}-$points at each $\mathbf{q}-$ points chosen along the high symmetry points. 

Within static theories such as DFT, equivalent to setting the DMFT self-energy is zero, $\chi^0(\mathbf{q},\omega)$ from Eq.\:\ref{eq:chiqw} is given as the bare susceptibility and  
its evaluation using Eq.\:\ref{eq:chi0} and Eq.\:\ref{eq:delta_form} reduces to the Lindhard formula for the  susceptibility:
\begin{eqnarray}
\chi^0(\mathbf{q},\omega) & = & \frac{(g\mu_B)^2}{2N}\sum_{\textbf{k},nm}  \sum_{\alpha\beta}\frac{f^m_{\mathbf{k+q}}-f^n_{\mathbf{k}}}{\omega+\epsilon^n_{\mathbf{k}}-\epsilon^m_{\mathbf{k+q}}+i0^+}\\
&&\cdot\langle\phi_{\alpha}^{\mathbf{k}}|\psi_n^{\mathbf{k}}\rangle\langle\psi_{n}^{\mathbf{k}}|\phi_{\beta}^{\mathbf{k}}\rangle
\langle\phi_{\beta}^{\mathbf{k+q}}|\psi_m^{\mathbf{k+q}}\rangle\langle\psi_{m}^{\mathbf{k+q}}|\phi_{\alpha}^{\mathbf{k+q}}\rangle,
\nonumber
\end{eqnarray}
where $\epsilon^n_{\mathbf{k}}$ is the eigenvalue of the DFT band at momentum $\mathbf{k}$ and the band index $n$. Therefore, it is expected that the $\chi^0(\mathbf{q})$ will be enhanced near the Fermi surface nesting $\mathbf{q}$ vector. This susceptibility calculation for multi-orbital systems based on DFT has been applied for various real materials~\cite{PhysRevB.73.205102,PhysRevB.90.115143,Chris_susceptibility}.


\section{Results and discussions}

We now present the results for the correlated electronic band structure and the Fermi surface, then relate them to the momentum dependent magnetic susceptibility of $M_{1/3}$NbS$_2$ ($M$=Co, Ni, and Fe) computed using DMFT.
In particular, we study the effect of the hole doping on the electronic structure and magnetism.

\subsection{Correlated electronic structure of Co$_{1/3}$NbS$_2$}

\begin{figure}[!ht]
\vspace{-0.3cm}
\includegraphics[width=0.8\linewidth]{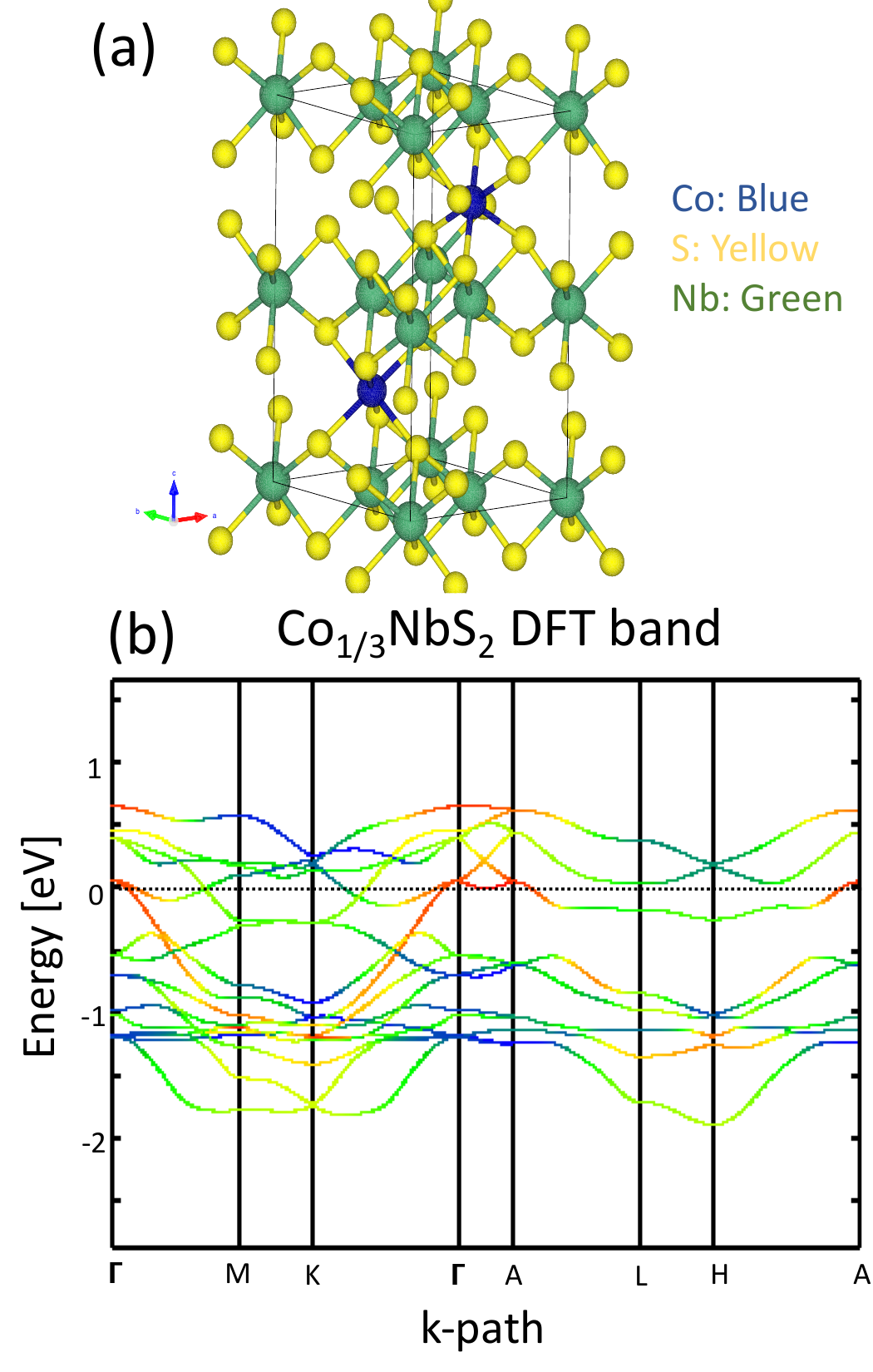}
\caption{ (a) The crystal structure of Co$_{1/3}$NbS$_2$, (b) The DFT band structure of Co$_{1/3}$NbS$_2$ projected to different orbital characters (red: Nb $d_{z^2}$ band, blue: Co $d$ band)
}
\label{fig:Co_DFT_Ak}
\end{figure}

\begin{figure}[!ht]
\includegraphics[width=0.8\linewidth]{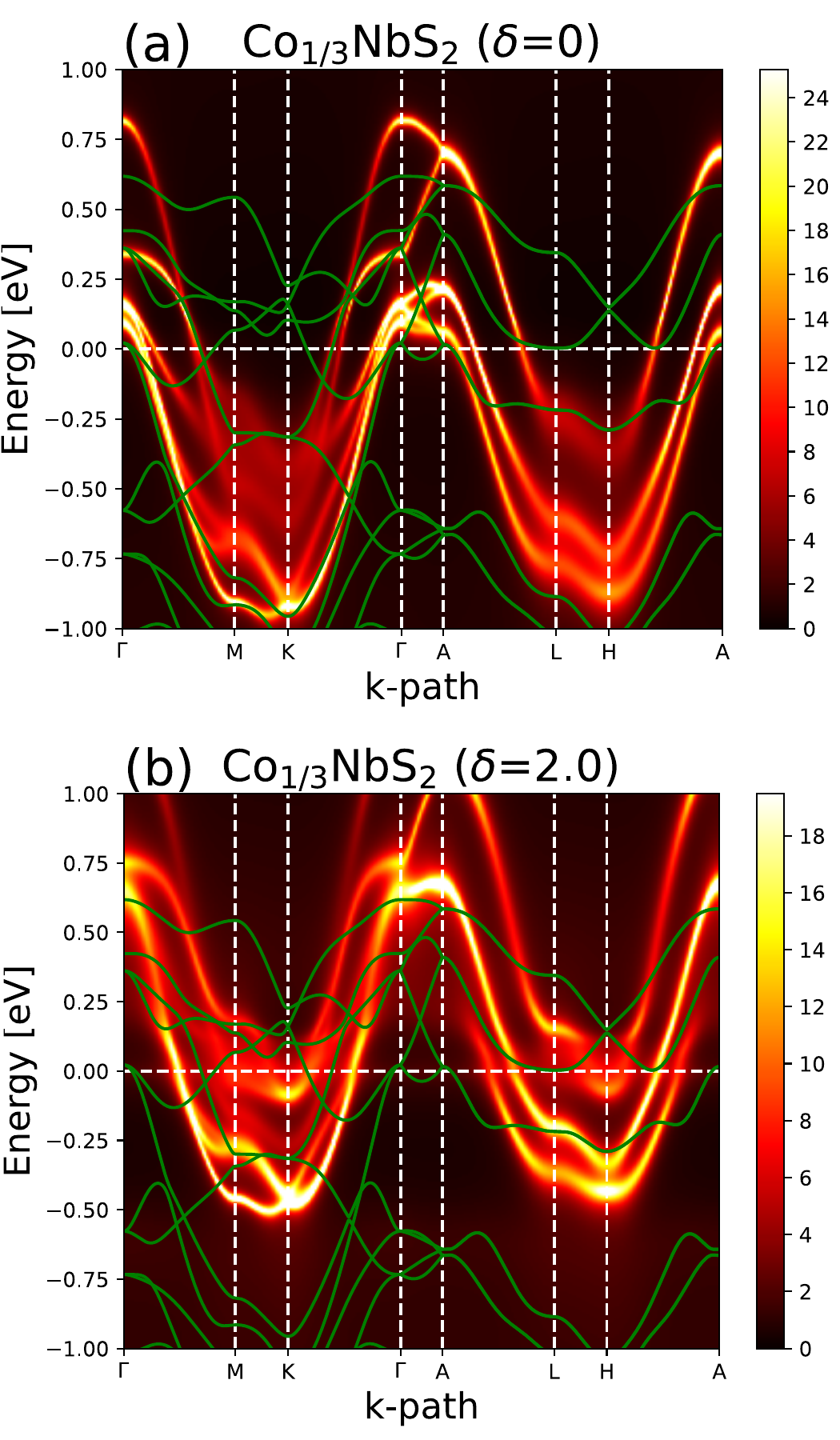}
\caption{ (a) The DMFT band structure of Co$_{1/3}$NbS$_2$, (b) The DMFT band structure upon the hole doping $\delta$=2.0 (two holes per CoNb$_3$S$_6$. The green lines represent the DFT band from Fig.\:\ref{fig:Co_DFT_Ak}(b).
}
\label{fig:Co_DMFT_Ak}
\end{figure}

The Co$_{1/3}$NbS$_2$ structure is formed by stacking NbS$_2$ layers with the intercalation of Co ions at two distinct Nb sites between the layers along the $c-$axis (see Fig.$\:$\ref{fig:Co_DFT_Ak}(a)).
To study the strong correlation effect of the Co $d$ orbitals on the band structure, we compare the DMFT spectral function $A(\mathbf{k},\omega)$  (Fig.\:\ref{fig:Co_DMFT_Ak}) to the DFT band structure in Fig.\:\ref{fig:Co_DFT_Ak}(b).
Here, we impose the paramagnetic spin symmetry (nonmagnetic state).
The DFT band (Fig.\:\ref{fig:Co_DFT_Ak}(b), the green thin lines in Fig.\:\ref{fig:Co_DMFT_Ak}) shows that the hole pocket near the $\Gamma$ and $A$ points are quite small and the band crossing near $K$ and $H$ points occur above the Fermi energy.
Fig.\:\ref{fig:Co_DFT_Ak}(b) shows the orbital characters of DFT band structure and the hole pockets are mostly the Nb $d_{z^2}$ character. The Co $d$ bands (blue color) are mostly located at $K$ and $H$ points above the Fermi energy with multiple degeneracy. 

The DMFT bands in Fig.\:\ref{fig:Co_DMFT_Ak}(a) show the strong modification of Co $d$ bands near $K$ and $H$ points as they are dressed by the DMFT self-energy. The location of Co $d$ bands is pushed below the Fermi energy and the spectra become much broader due to the large self-energy effect. The Nb $d_{z^2}$ bands still show the quasi-particle dispersion along with the feature of the well-defined Fermi surface.
We also study the doping effect by changing the total number of  valence electrons within the DMFT calculation and computing the corresponding band structure.
The hole-doping effect on the DMFT bands in Fig.\:\ref{fig:Co_DMFT_Ak}(b) shows the crossing of Co $d$ bands near the Fermi energy near $K$ and $H$ points due to the up-shift of bands.
Upon the hole doping, the size of the hole pocket near the $\Gamma$ point increases as the Nb $d_{z^2}$ band shifts upward.
An important feature of the hole-doping on the Co $3d$ band structure is the appearance of the small broad electron pocket at the $K$ point, which is consistent with the ARPES measurement \cite{PhysRevB.105.L121102}.

In Co$_{1/3}$NbS$_2$, our DMFT calculation shows that the occupancy of the Co $d$ orbital is close to 7.0 meaning that the Co ion has the valence state of $2+$. 
Since the S ion is the $2-$ state, the valence state of Nb is close to $(10/3)+$.
Therefore, the Nb ion has the occupancy of $4d^{1.67}$, which is larger than the $4d^1$ one (Nb$^{4+}$) of the pure NbS$_2$ layer.
In other words, the hole pocket at the $\Gamma$ point of the pure NbS$_2$ is expected to be larger than that of Co$_{1/3}$NbS$_2$ (Fig.\:\ref{fig:Co_DMFT_Ak}(a)).
Our hole doping effect of $\delta$=2.0 (two holes per CoNb$_3$S$_6$) means that 2/3 holes are mostly doped to the Nb ions since the DMFT occupancy of the Co $d$ orbitals still remains close to 7.0. 
As a result, Co$_{1/3}$NbS$_2$ has the strong hybridization between Co $3d$ and Nb $4d_{z^2}$ orbitals resulting the doped holes residing mostly on the Nb $d_{z^2}$ orbital.
Therefore, this hole doping effect mainly affects the size the hole pocket near the $\Gamma$ point.

\begin{figure*}[!ht]
\includegraphics[width=1\linewidth]{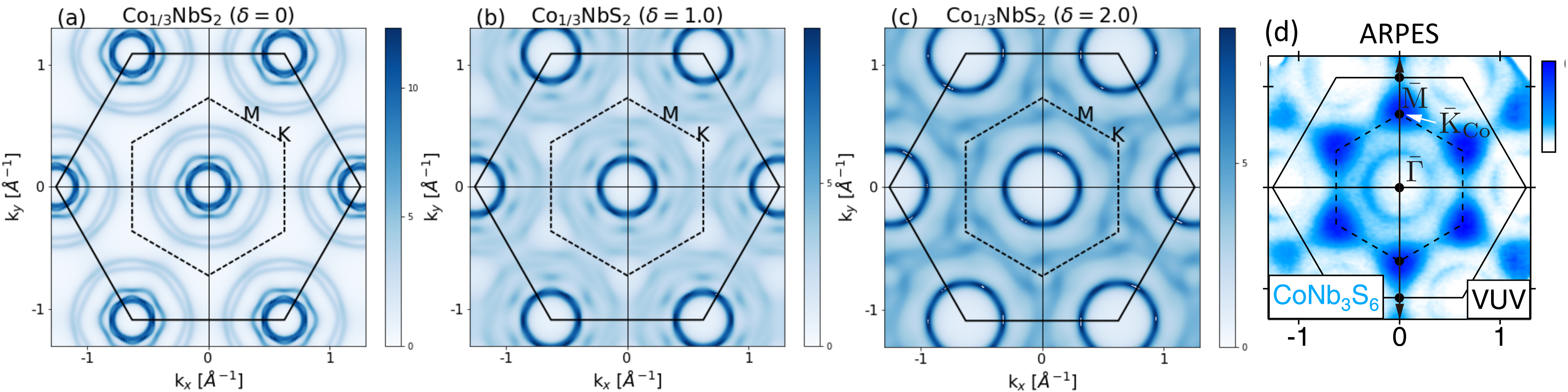}
\caption{The DMFT Fermi surface for Co$_{1/3}$NbS$_2$ as a function of hole-doping (a) $\delta$=0, (b) 1.0, and (c) 2.0. The hole-doping effect increases the size of the hole pocket centered at the $\Gamma$ point. The calculated Fermi surface is consistent with the (d) experimental ARPES measurement~\cite{PhysRevB.105.L121102} when $\delta$=2.0. The dashed line represents the BZ of Co$_{1/3}$NbS$_2$ and the solid line shows the BZ of NbS$_2$. Note that the $M$ point in (d) is defined in the solid-line BZ, while our $M$ points are defined in the dashed-line BZ.
}
\label{fig:Co_FS}
\end{figure*}

Fig.\:\ref{fig:Co_FS} shows the Fermi surface of Co$_{1/3}$NbS$_2$ computed using DFT+DMFT at different hole doping $\delta$ values ($\delta$= the number of holes per Co ion).
At $\delta$=0, the DMFT hole-pocket centered at the $\Gamma$ point has a circular shape similarly as measured in ARPES although its size is smaller than the ARPES one.
The outer larger pocket has mostly Co $d$ orbital character and exhibits much weaker intensity due to the large scattering rate ($Im\Sigma(\omega)$).
Upon hole-doping, the smaller hole-pocket gets larger in size, comparable to the ARPES measurement and the Co $d$ states move closer to the $K$ point in the BZ. 
This broad Co $d$ spectra near the $K$ point is also captured in ARPES.
Our DMFT Fermi surface calculation shows that the hole doping of $\delta$=2.0 makes the size of Nb hole pocket in the Fermi surface similar to the ARPES measurement.

Since the ARPES measurement is sensitive to the surface state of possibly NbS$_2$ termination layers, the electronic structure will have the hole-doping effect due to the Co ion deficiency.
In the bulk, Co or Nb vacancies can induce the effect of hole-dopings similarly to the surface state. Our DMFT calculation shows that this doping effect can tune the size of hole pocket significantly, possibly affecting the magnetic properties. Previous experimental study also shows that the AHE  of Co$_{1/3}$NbS$_{2-x}$ can be dramatically changed by the S deficiency, which can lead to the similar  doping effect~\cite{PhysRevB.105.L121102}.

 
\subsection{Co$_{1/3}$NbS$_2$ magnetic susceptibility}

\begin{figure}[!ht]
\includegraphics[width=\linewidth]{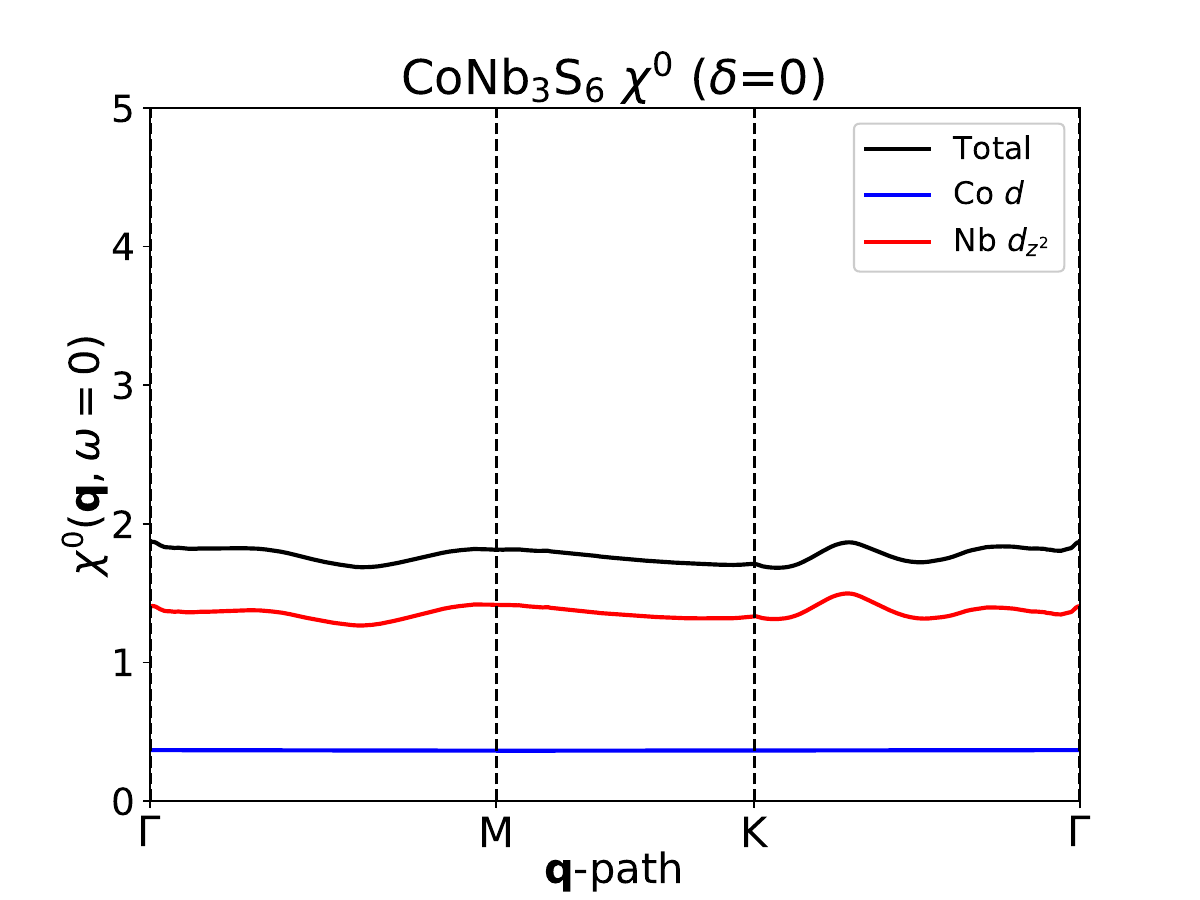}\\
\includegraphics[width=\linewidth]{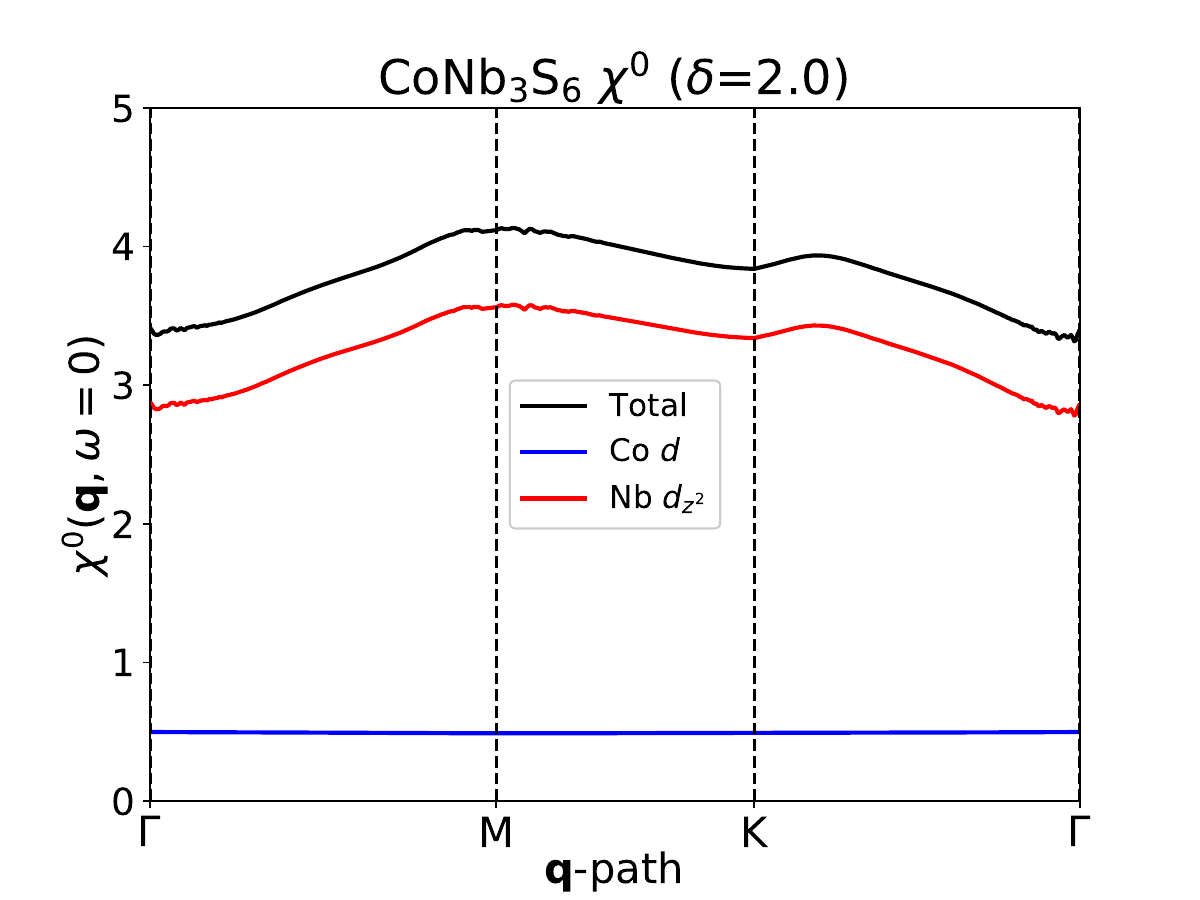}
\caption{The polarizability $\chi^0$ (Eq.\:\ref{eq:chi0}) for Co$_{1/3}$NbS$_2$ computed at different hole doping $\delta$, i.e. $\delta$=0 (top panel) and $\delta$=2.0 (bottom panel). 
}
\label{fig:Co_chi0}
\end{figure}

Our previous DFT calculation on Co$_{1/3}$NbS$_2$ shows that a $3q-$type magnetic structure is energetically stable and may  be responsible for the large observed anomalous Hall current in this material.
This particular $3q-$type spin structure corresponds to the non-coplanar spins arrangement with the four spins within the magnetic unit cell of one Co layer pointing towards four vertices of a tetrahedron in the spin space.
Such unusual spin configuration can be stabilized  in a triangular lattice due to the Fermi surface nesting of itinerant electrons~\cite{martin_scalar_2008}. For every triangular  plaquette of the spin lattice,  scalar spin chirality, $\chi_{ijk} = {\bf S}_i\cdot [{\bf S}_j\times {\bf S}_k]$ is constant, corresponding  to a uniform Berry flux per plaquette.
The modulation vector $\mathbf{q}$ of this $3q-$type spin structure is half of the reciprocal lattice vectors of the primitive unit cell, i.e., the high-symmetry $M$ points in the Brillouine zone. 
Early  neutron scattering experiment \cite{parkin_magnetic_1983} showed indeed the scattering peak  at $\mathbf{q}=(1/2,0,0)$ ($M$ point). However,  confirming the existence of the non-coplanar 3$q$ state requires more complex polarized neutron scattering experiments, as has been successfully  realized in some cases ~\cite{doi:10.1126/sciadv.aau3402,NaturePhysics_Takagi}.

Neutron scattering measures magnetic susceptibility.
We here compute the momentum-dependent magnetic susceptibility, $\chi(\mathbf{q},\omega=0)$ of Co$_{1/3}$NbS$_2$ at different hole-doping $\delta$ values to understand the spin modulation vector $\mathbf{q}$ of the leading magnetic instability and its relation to the correlated electronic structure.
We first compute the real part of the bare susceptibility $\chi^0$ using DMFT (Eq.\:\ref{eq:chi0}) at $\delta=0$ and 2.0, as shown in Fig.\:\ref{fig:Co_chi0}. 
In both doping levels, the Co bare spin susceptibility ($\chi^0$) obtained from DMFT shows very weak momentum dependence due to the the strongly localized nature of Co $d$ orbitals.
This can be understood from the one-particle DMFT spectra of Co $d$ bands showing no clear evidence of quasi-particle peaks near the Fermi energy, rather very broad feature of band dispersion without much dependence on momenta.
Unlike the Co $d$ orbitals, 
the Nb $d_{z^2}$ orbitals show a rather strong momentum dependence of the susceptibility at $\delta=2.0$ due to their itinerant nature near the Fermi energy. 
The contribution of the Nb $d_{z^2}$ orbitals to the susceptibility $\chi^0$ is larger than that of the Co $d$ orbitals and
depends sensitively on momentum and the the doping levels (see Fig.\:\ref{fig:Co_chi0}).


We argue that the leading modulation vector $\mathbf{q}$ of $\chi^0$ in Co$_{1/3}$NbS$_2$ can be mostly determined by the Fermi surface  momentum ($2k_F$) of the hole pocket centered at the $\Gamma$ point. As shown in Fig.\:\ref{fig:Co_FS}, the size of the hole pocket can be sensitively dependent on the hole doping levels and it is consistent with the ARPES measurement at $\delta=2.0$.
At $\delta$=0, the Nb $d_{z^2}$ orbital contribution to $\chi^0$ has no clear momentum dependence although the spectra near the $\Gamma$ point are slightly larger than those at other momenta. This is because the smallest hole-pocket near the $\Gamma$ point has the largest spectral weight while the other Fermi surfaces have much smaller spectral weights. 
As the hole doping increases ($\delta$=2.0), the size of the hole pocket 
near the $\Gamma$ point also increases and the contribution of the Nb $d_{z^2}$ orbital to the susceptibility favors the modulation $\mathbf{q}$ vector at the high-symmetry $M$ point, which is close to the Nb Fermi surface momentum (2$k_F$). 
The susceptibility near the $K$ point also shows the enhanced peak height at $|\mathbf{q}|\simeq 2k_F$ although the $M$ point shows the maximum peak height.
We note that  the peak heights of the susceptibility  depend on the effect of the form factor in Eq.\:\ref{eq:form_final}  -- the peak height at the $M$ point becomes much closer to that at the $K$ point if the form factor is simplified using Eq.\:\ref{eq:delta_form} (see Appendix).


\begin{figure}[!ht]
\includegraphics[width=\linewidth]{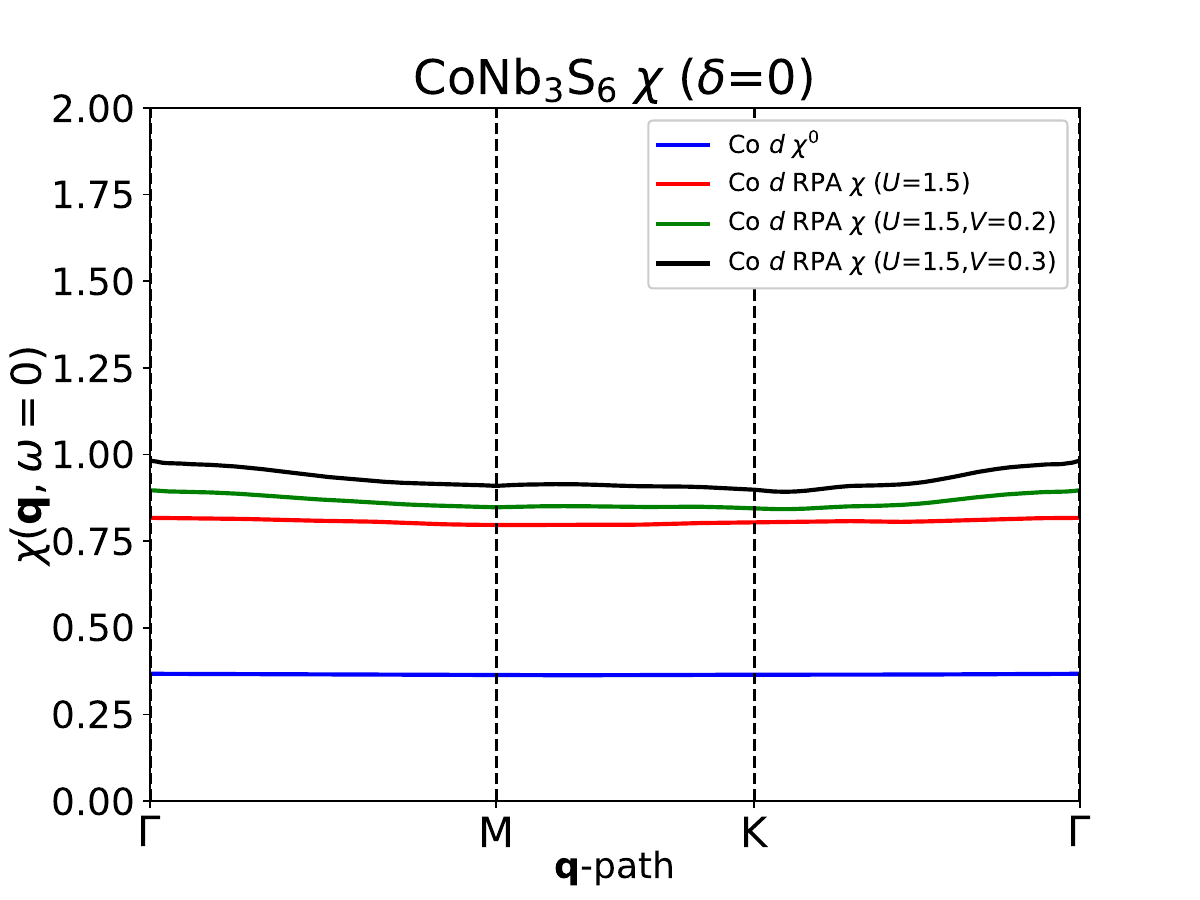}\\
\includegraphics[width=\linewidth]{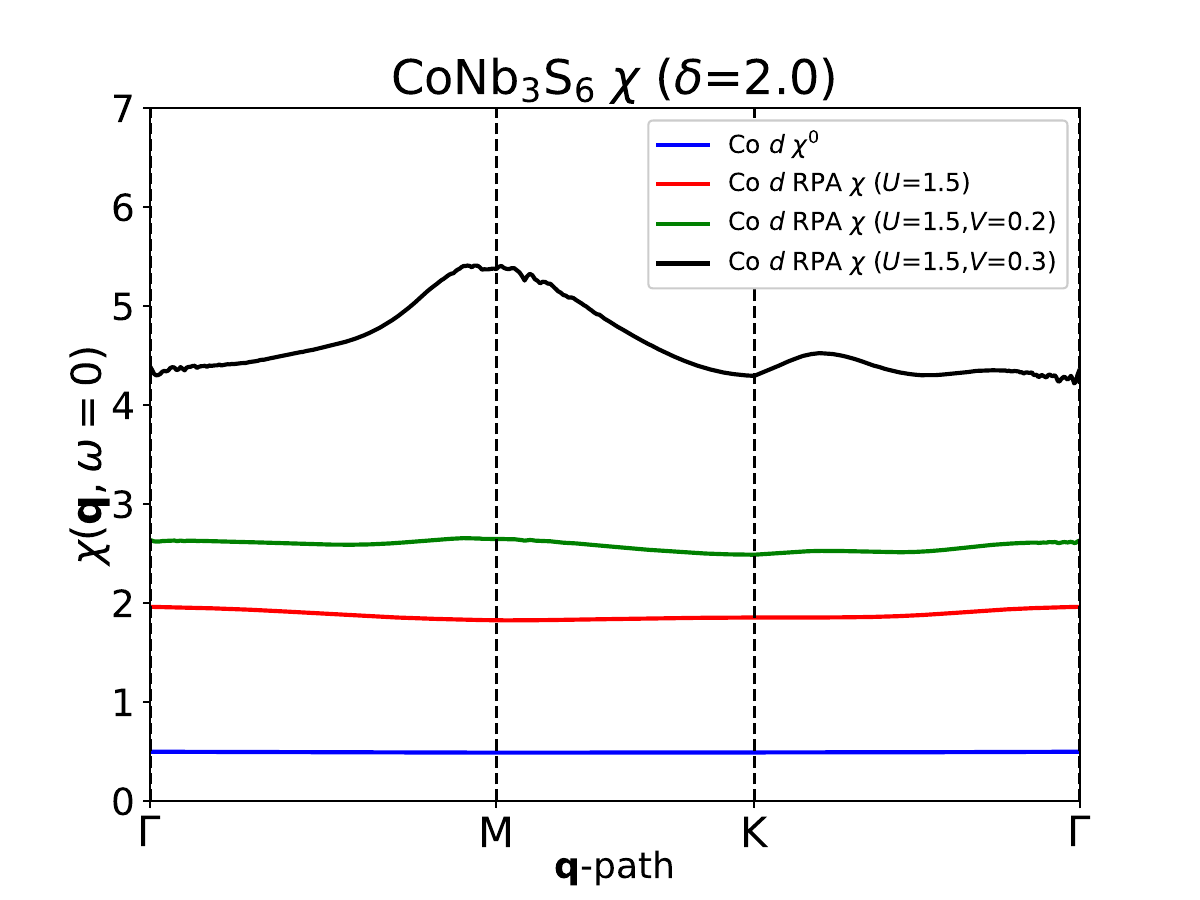}
\caption{The magnetic susceptibility $\chi$ for Co$_{1/3}$NbS$_2$ including the RPA-type interaction at different hole-dopings $\delta=0$ (top panel) and $\delta=2.0$ (bottom panel). 
}
\label{fig:Co_chi}
\end{figure}

While the bare magnetic susceptibility $\chi^0$ of Co $d$ orbitals does not have any significant momentum dependence due to the strongly localized nature of the band structure, the full magnetic susceptibility 
including the interaction effect shows the preference for a particular momentum suggesting the long-range spin ordering of Co $d$ spins coupled via the RKKY interaction mediated by the itinerant Nb $d_{z^2}$ bands. 
It turns out that the inter-site interaction $\bar{V}$ plays an important role in mediating the RKKY interaction. Our RPA susceptibility calculation shows that the local interaction $\bar{U}$ enhances the absolute value of $\chi(\mathbf{q})$ while retaining the weak momentum dependence. The increase of $\bar{V}$ results in the momentum dependence of $\chi(\mathbf{q})$, which is peaked at the $M$ point (the same peak position as $\chi^0$) for the RPA susceptibility (see Fig.\:\ref{fig:Co_chi}). 

We find that the $\chi$ diverges at the $M$ point near $\bar{U}$=1.7eV and $\bar{V}$=0.3eV, supporting the occurrence of the magnetic instability.
While this is a direct way to study the instability from the susceptibility, one can also further analyze the different band contributions to the magnetic instability by decomposing the product of the $\chi^0$ and the $\Gamma^{irr}$ matrices while solving the Bethe-Salpeter equation in Eq.\:\ref{eq:BSE}~\cite{Chris_susceptibility}.
Again, the reasonable range of the RPA $\bar{U}$ should be much smaller than the Hubbard $U$ used in DMFT since it does not account for the orbital and dynamical screening process.
The screened inter-site $\bar{V}$ also should be much smaller than the on-site $\bar{U}$ value.
While determining $\bar{U}$ and $\bar{V}$ quantitatively will be a complicated task, we find that the qualitative feature of the magnetic susceptiblity (i.e., the momentum dependence) does not vary depending on $\bar{U}$ and $\bar{V}$ values.

\begin{figure}[!ht]
\includegraphics[width=0.95\linewidth]{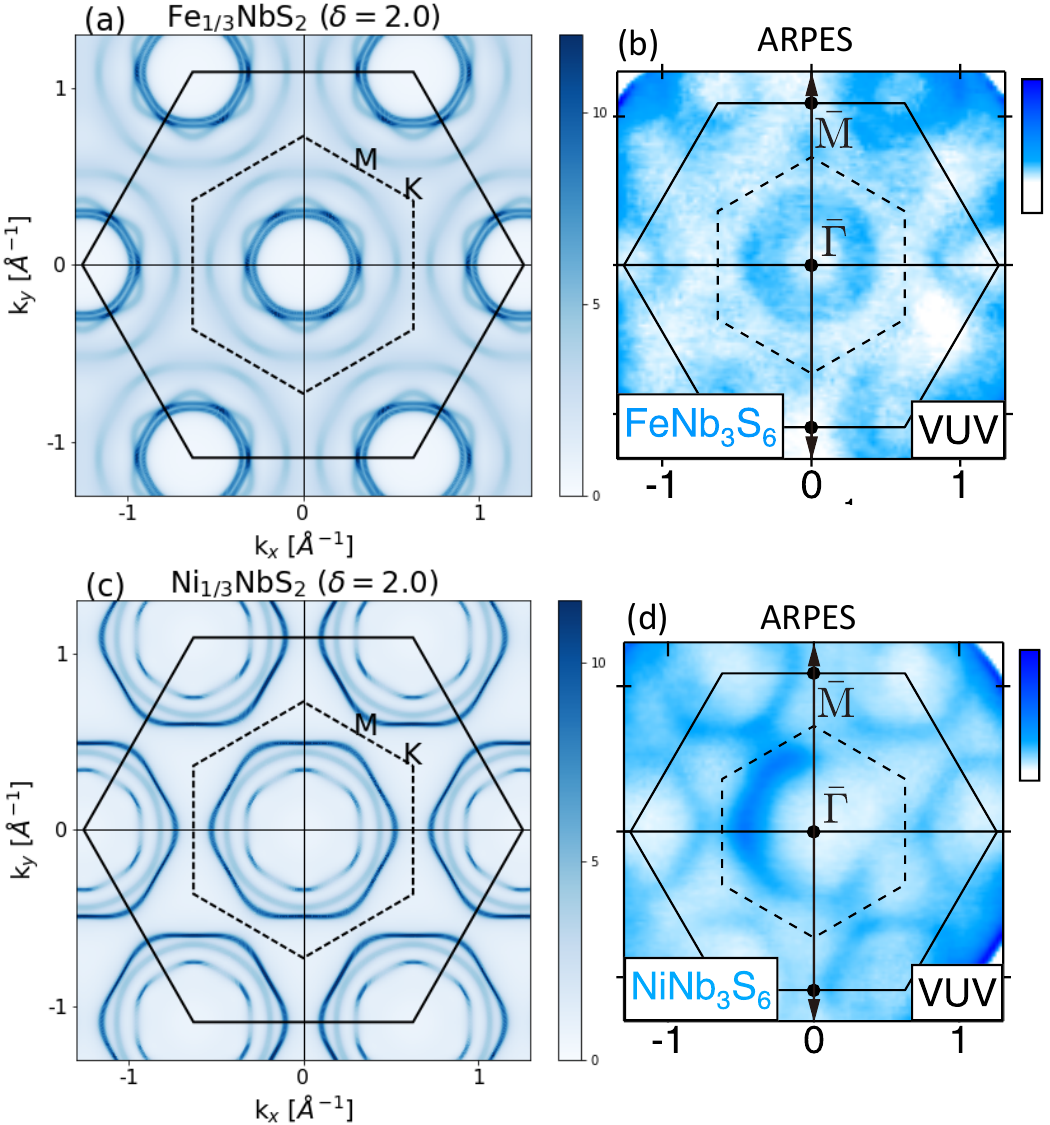}
\caption{The DMFT Fermi surfaces (left panels) for $M_{1/3}$NbS$_2$ ($M$= (a) Fe (top panel) and (c) Ni (bottom panel)) at the hole doping of $\delta$= 2.0. The  experimental ARPES~\cite{PhysRevB.105.L121102} measurements are also compared (right panels). 
}
\label{fig:Fe_FS}
\end{figure}

\subsection{Electronic structure of Fe$_{1/3}$NbS$_2$ and Ni$_{1/3}$NbS$_2$}

While Co$_{1/3}$NbS$_2$ shows  large anomalous Hall effect likely  originating from the non-coplanar spin structure, such effects have not been seen experimentally for Fe$_{1/3}$NbS$_2$ and Ni$_{1/3}$NbS$_2$. 
Although our previous DFT calculation showed that both Fe$_{1/3}$NbS$_2$ and Ni$_{1/3}$NbS$_2$ can favor the non-coplanar $3q$ spin structure energetically, 
the ground-state magnetic state has been studied only for a small number of possible  $\mathbf{q}$ vectors allowed within a supercell.
Therefore, it is plausible that the leading magnetic $\mathbf{q}-$vector can vary depending on the intercalated transition metal ions due to the electronic structure change.  
Here, we compute the magnetic susceptibility and the Fermi surface of Fe$_{1/3}$NbS$_2$ and Ni$_{1/3}$NbS$_2$ at $\delta=2.0$, similarly as the Co$_{1/3}$NbS$_2$ case. 

We find that the hole doping effects in Fe$_{1/3}$NbS$_2$ and Ni$_{1/3}$NbS$_2$ can be quite different.
Fig.\:\ref{fig:Fe_FS} shows that the DMFT Fermi surface of Fe$_{1/3}$NbS$_2$ has a slightly smaller hole pocket compared to the Co$_{1/3}$NbS$_2$ one at the same hole doping ($\delta$=2.0).
Our DMFT calculation shows that the hole doping induces the change of the Fe $d$ occupancy as the Fe valence state becomes close to Fe$^{2.3+}$. 
In Fe$_{1/3}$NbS$_2$, the hole doping  mostly affects the Fe states near the Fermi energy since the hybridization between the Fe $3d$ and Nb $4d$ orbitals is rather weak.
As a result, the size of the Nb hole pocket in Fe$_{1/3}$NbS$_2$ is less sensitive to dopings while it becomes larger upon hole doping for the Co$_{1/3}$NbS$_2$ case.
Moreover, the Fe $d$ character becomes much weaker and not visible near the $K$ point, as also seen in the ARPES data~\cite{PhysRevB.105.L121102}.
In Ni$_{1/3}$NbS$_2$, the DMFT valence of the Ni ion is still close to Ni$^{2+}$ and the Ni band is located lower than the Co or Fe bands in the other materials.
This means that the Ni band has the negative charge-transfer effect, similarly to what is observed  in some other rare-earth nickelates~\cite{nickelate}.
As a result, the hole doping mostly affects the Nb hole states and the Nb hole pocket in Ni$_{1/3}$NbS$_2$ becomes the largest among three materials.

\begin{figure}[!ht]
\includegraphics[width=\linewidth]{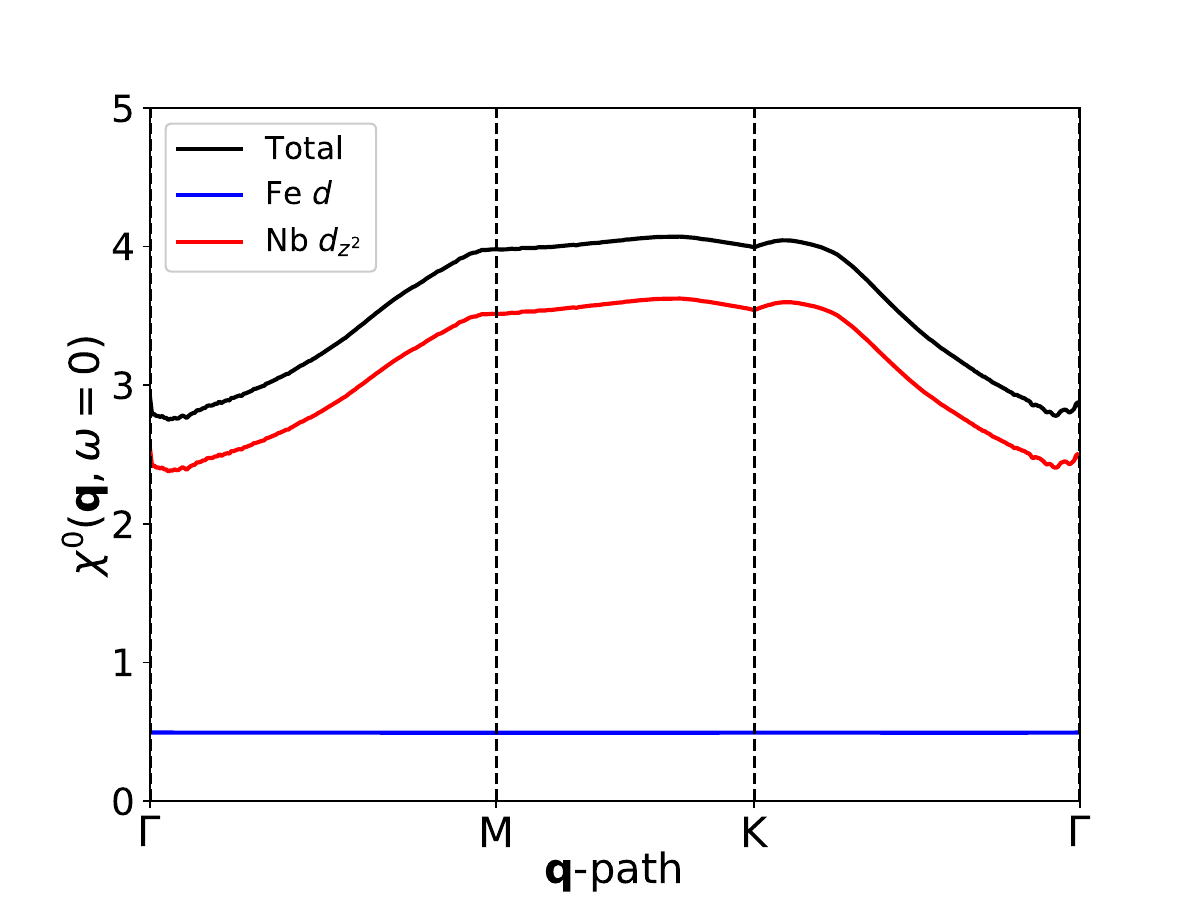}\\
\includegraphics[width=\linewidth]{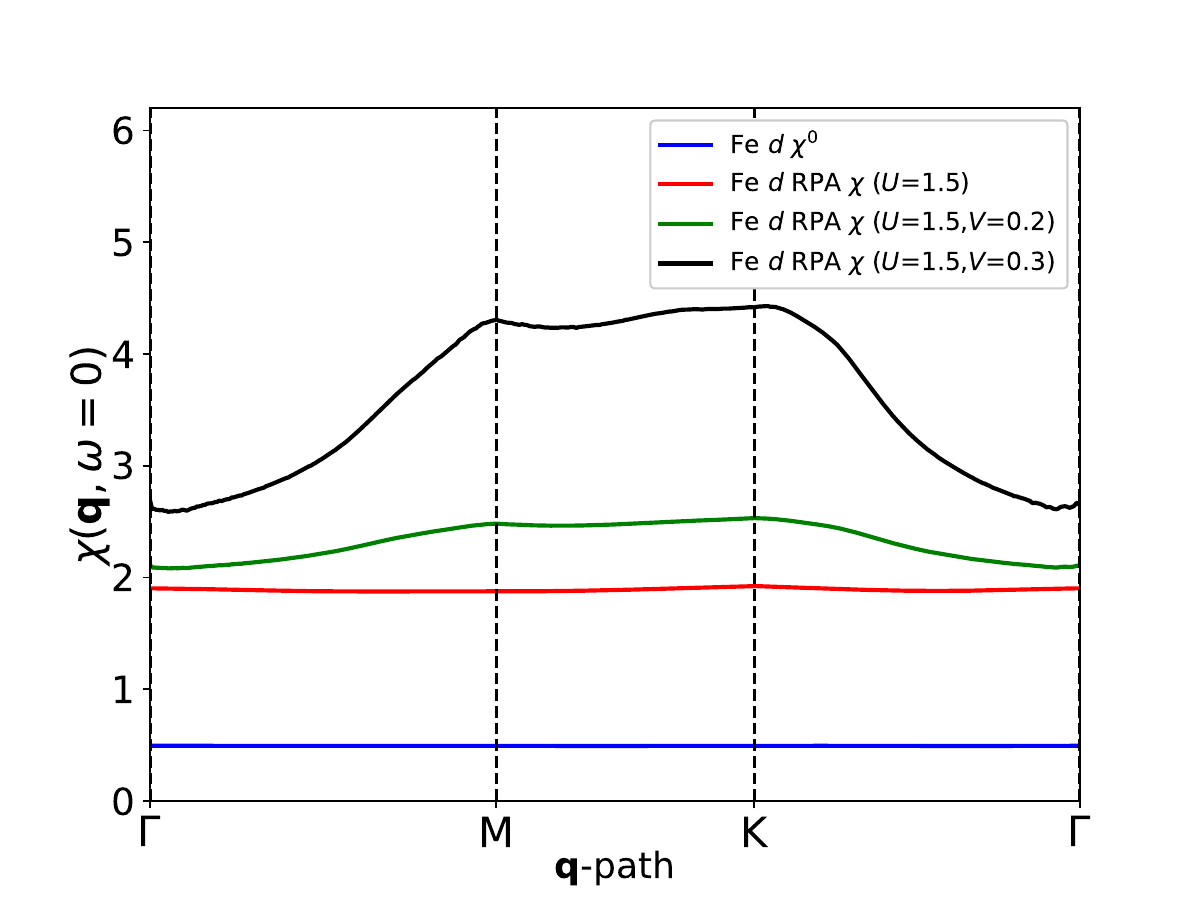}
\caption{The magnetic susceptibility for Fe$_{1/3}$NbS$_2$, the polarizability $\chi^0$ (top panel) and the RPA $\chi$ (bottom panel) 
}
\label{fig:Fe_chi}
\end{figure}

\begin{figure}[!ht]
\includegraphics[width=\linewidth]{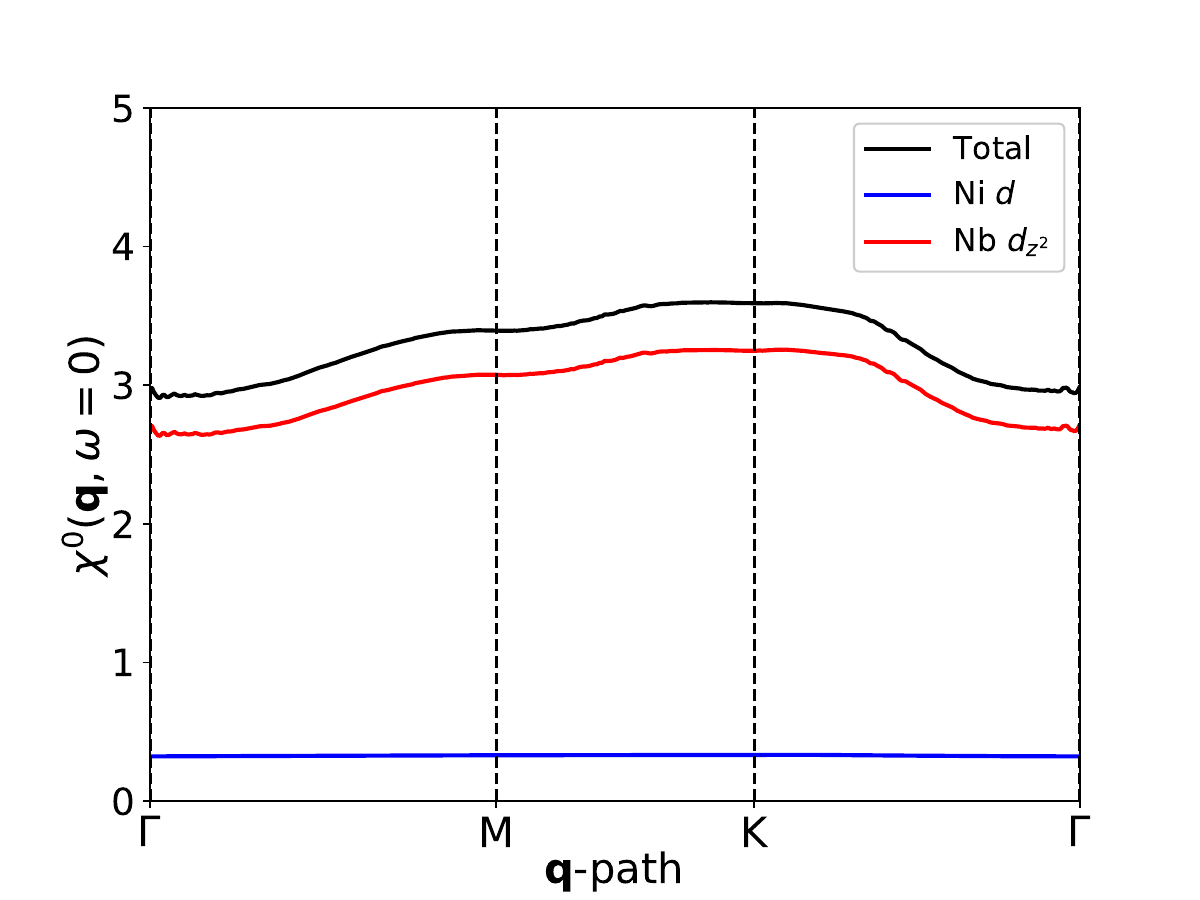}\\
\includegraphics[width=\linewidth]{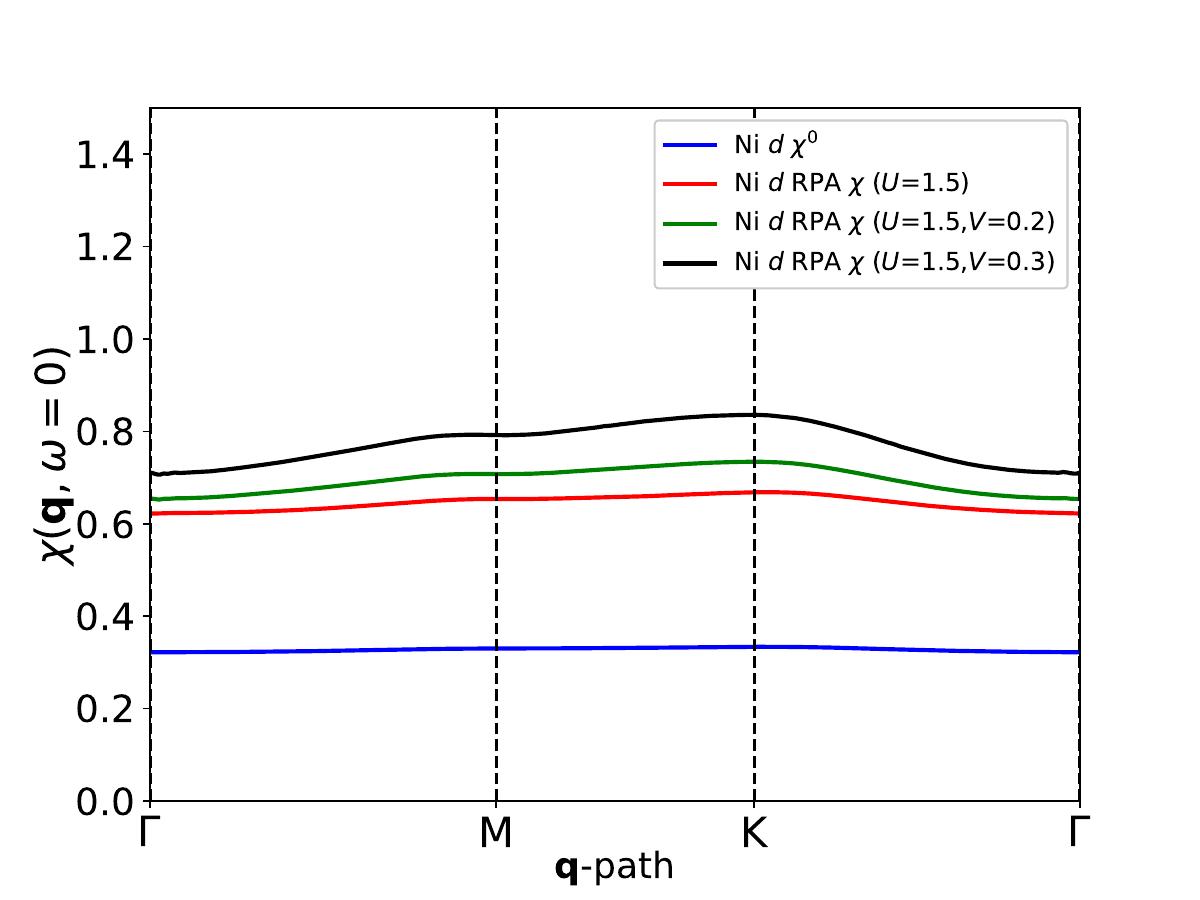}
\caption{The magnetic susceptibility for Ni$_{1/3}$NbS$_2$, the polarizability $\chi^0$ (top panel) and the RPA $\chi$ (bottom panel) 
}
\label{fig:Ni_chi}
\end{figure}

\subsection{The magnetic susceptibility of Fe$_{1/3}$NbS$_2$ and Ni$_{1/3}$NbS$_2$ }

These notable variations of the DMFT Fermi surface due to the intercalation by different transition metals can also change the leading magnetic susceptibility momentum $\mathbf{q}$.
In Fe$_{1/3}$NbS$_2$, the susceptibility $\mathbf{q}$ peaks near $M$ and $K$ points are almost degenerate as the the Fermi momentum $2k_F$ of the hole pocket in the Fermi surface gets closer to both $M$ and $K$ points. 
Similar to the Co$_{1/3}$NbS$_2$ case, the inter-site interaction $V$ strongly enhances the momentum dependence of the Fe $d$ susceptibility.
In Ni$_{1/3}$NbS$_2$, the susceptibility peak is slightly higher at the $K$ point as the leading modulation vector, 
while the momentum dependence of the susceptibility is the weakest among three materials. 
This is because 
the $2k_F$ of the hole pocket in Ni$_{1/3}$NbS$_2$ is much larger than the high symmetry points and, as a result, the susceptibility does not show the dominant momentum peak.

In both cases of Fe$_{1/3}$NbS$_2$ and Ni$_{1/3}$NbS$_2$, the modulation factor (Eq.\:\ref{eq:form}) in the susceptibility can change the $\chi(\mathbf{q})$ profile. 
Without the factor, both susceptibilities enhance the peak near the $K$ point (see Appendix).
It is possible that both Fe$_{1/3}$NbS$_2$ and Ni$_{1/3}$NbS$_2$ can have different magnetic $\mathbf{q}$ instability compared to the Co$_{1/3}$NbS$_2$ case.
Finally, the spectral weight of the susceptibility is the smallest for Ni$_{1/3}$NbS$_2$ as the static magnetic moment of the Ni ion is the smallest compared to the other materials~\cite{anzenhofer}.
 
\section{Conclusion}

We studied the magnetic susceptibility and the correlated electronic structure of $M_{1/3}$NbS$_2$ ($M$= Co, Fe, and Ni) using DMFT to treat the strong correlation effect of transition metal ions.
Our DMFT band structure and the Fermi surface calculations upon hole-doping are consistent with the ARPES measurements \cite{PhysRevB.105.L121102} of these compounds.
The size of the hole pocket centered at the $\Gamma$ point is comparable to the ARPES data for all compounds and the appearance of the electron pocket at the $K$ point in Co$_{1/3}$NbS$_2$ is correctly captured.
Due to the strong hybridization between Co/Ni $3d$ orbitals and Nb $4d$ orbitals, the doped holes mostly change the Nb valence states and the size of the hole pocket centered at the $\Gamma$ point can be tuned upon the hole doping.
In Fe$_{1/3}$NbS$_2$, the hole-doping effect changes mostly the Fe valence state.

We also show that the spin susceptibility $\chi(\mathbf{q})$ calculation using DMFT can help to identify the  momentum $\mathbf{q}$ of the leading magnetic instability in strongly correlated materials. 
This method allows to avoid the need for constructing a large supercell to study the magnetic instability at the arbitrary momentum $\mathbf{q}$ vector.
While the spin susceptibility of Co$_{1/3}$NbS$_2$ is peaked at $\mathbf{q}=(1/2,0,0)$ (the $M$ point), which is consistent with the $3q$-type non-coplanar spin structure, the maximum peak positions change to the $K$ point for Fe$_{1/3}$NbS$_2$ and Ni$_{1/3}$NbS$_2$.
This suggests that the magnetic ground state of these two compounds will be distinct from that of Co$_{1/3}$NbS$_2$.

\section*{Acknowledgement}
We would like to thanks Mike Norman and Chris Lane for useful discussions.
This work was supported by the Materials Sciences and Engineering Division, Basic Energy Sciences, Office of Science, US Department of Energy. We gratefully acknowledge the computing resources provided on Bebop, a high-performance computing cluster operated by the Laboratory Computing Resource Center at Argonne National Laboratory.

\bibliography{main}

\begin{thebibliography}{25}%
\makeatletter
\providecommand \@ifxundefined [1]{%
 \@ifx{#1\undefined}
}%
\providecommand \@ifnum [1]{%
 \ifnum #1\expandafter \@firstoftwo
 \else \expandafter \@secondoftwo
 \fi
}%
\providecommand \@ifx [1]{%
 \ifx #1\expandafter \@firstoftwo
 \else \expandafter \@secondoftwo
 \fi
}%
\providecommand \natexlab [1]{#1}%
\providecommand \enquote  [1]{``#1''}%
\providecommand \bibnamefont  [1]{#1}%
\providecommand \bibfnamefont [1]{#1}%
\providecommand \citenamefont [1]{#1}%
\providecommand \href@noop [0]{\@secondoftwo}%
\providecommand \href [0]{\begingroup \@sanitize@url \@href}%
\providecommand \@href[1]{\@@startlink{#1}\@@href}%
\providecommand \@@href[1]{\endgroup#1\@@endlink}%
\providecommand \@sanitize@url [0]{\catcode `\\12\catcode `\$12\catcode
  `\&12\catcode `\#12\catcode `\^12\catcode `\_12\catcode `\%12\relax}%
\providecommand \@@startlink[1]{}%
\providecommand \@@endlink[0]{}%
\providecommand \url  [0]{\begingroup\@sanitize@url \@url }%
\providecommand \@url [1]{\endgroup\@href {#1}{\urlprefix }}%
\providecommand \urlprefix  [0]{URL }%
\providecommand \Eprint [0]{\href }%
\providecommand \doibase [0]{http://dx.doi.org/}%
\providecommand \selectlanguage [0]{\@gobble}%
\providecommand \bibinfo  [0]{\@secondoftwo}%
\providecommand \bibfield  [0]{\@secondoftwo}%
\providecommand \translation [1]{[#1]}%
\providecommand \BibitemOpen [0]{}%
\providecommand \bibitemStop [0]{}%
\providecommand \bibitemNoStop [0]{.\EOS\space}%
\providecommand \EOS [0]{\spacefactor3000\relax}%
\providecommand \BibitemShut  [1]{\csname bibitem#1\endcsname}%
\let\auto@bib@innerbib\@empty
\bibitem [{\citenamefont {Ghimire}\ \emph {et~al.}(2018)\citenamefont
  {Ghimire}, \citenamefont {Botana}, \citenamefont {Jiang}, \citenamefont
  {Zhang}, \citenamefont {Chen},\ and\ \citenamefont
  {Mitchell}}]{ghimire_large_2018}%
  \BibitemOpen
  \bibfield  {author} {\bibinfo {author} {\bibfnamefont {N.~J.}\ \bibnamefont
  {Ghimire}}, \bibinfo {author} {\bibfnamefont {A.~S.}\ \bibnamefont {Botana}},
  \bibinfo {author} {\bibfnamefont {J.~S.}\ \bibnamefont {Jiang}}, \bibinfo
  {author} {\bibfnamefont {J.}~\bibnamefont {Zhang}}, \bibinfo {author}
  {\bibfnamefont {Y.-S.}\ \bibnamefont {Chen}}, \ and\ \bibinfo {author}
  {\bibfnamefont {J.~F.}\ \bibnamefont {Mitchell}},\ }\href
  {https://www.nature.com/articles/s41467-018-05756-7} {\bibfield  {journal}
  {\bibinfo  {journal} {Nature Communications}\ }\textbf {\bibinfo {volume}
  {9}},\ \bibinfo {pages} {3280} (\bibinfo {year} {2018})}\BibitemShut
  {NoStop}%
\bibitem [{\citenamefont {Tenasini}\ \emph {et~al.}(2020)\citenamefont
  {Tenasini}, \citenamefont {Martino}, \citenamefont {Ubrig}, \citenamefont
  {Ghimire}, \citenamefont {Berger}, \citenamefont {Zaharko}, \citenamefont
  {Wu}, \citenamefont {Mitchell}, \citenamefont {Martin}, \citenamefont
  {Forr\'o},\ and\ \citenamefont {Morpurgo}}]{PhysRevResearch.2.023051}%
  \BibitemOpen
  \bibfield  {author} {\bibinfo {author} {\bibfnamefont {G.}~\bibnamefont
  {Tenasini}}, \bibinfo {author} {\bibfnamefont {E.}~\bibnamefont {Martino}},
  \bibinfo {author} {\bibfnamefont {N.}~\bibnamefont {Ubrig}}, \bibinfo
  {author} {\bibfnamefont {N.~J.}\ \bibnamefont {Ghimire}}, \bibinfo {author}
  {\bibfnamefont {H.}~\bibnamefont {Berger}}, \bibinfo {author} {\bibfnamefont
  {O.}~\bibnamefont {Zaharko}}, \bibinfo {author} {\bibfnamefont
  {F.}~\bibnamefont {Wu}}, \bibinfo {author} {\bibfnamefont {J.~F.}\
  \bibnamefont {Mitchell}}, \bibinfo {author} {\bibfnamefont {I.}~\bibnamefont
  {Martin}}, \bibinfo {author} {\bibfnamefont {L.}~\bibnamefont {Forr\'o}}, \
  and\ \bibinfo {author} {\bibfnamefont {A.~F.}\ \bibnamefont {Morpurgo}},\
  }\href {\doibase 10.1103/PhysRevResearch.2.023051} {\bibfield  {journal}
  {\bibinfo  {journal} {Phys. Rev. Res.}\ }\textbf {\bibinfo {volume} {2}},\
  \bibinfo {pages} {023051} (\bibinfo {year} {2020})}\BibitemShut {NoStop}%
\bibitem [{\citenamefont {Nagaosa}\ \emph {et~al.}(2010)\citenamefont
  {Nagaosa}, \citenamefont {Sinova}, \citenamefont {Onoda}, \citenamefont
  {MacDonald},\ and\ \citenamefont {Ong}}]{RevModPhys.82.1539}%
  \BibitemOpen
  \bibfield  {author} {\bibinfo {author} {\bibfnamefont {N.}~\bibnamefont
  {Nagaosa}}, \bibinfo {author} {\bibfnamefont {J.}~\bibnamefont {Sinova}},
  \bibinfo {author} {\bibfnamefont {S.}~\bibnamefont {Onoda}}, \bibinfo
  {author} {\bibfnamefont {A.~H.}\ \bibnamefont {MacDonald}}, \ and\ \bibinfo
  {author} {\bibfnamefont {N.~P.}\ \bibnamefont {Ong}},\ }\href {\doibase
  10.1103/RevModPhys.82.1539} {\bibfield  {journal} {\bibinfo  {journal} {Rev.
  Mod. Phys.}\ }\textbf {\bibinfo {volume} {82}},\ \bibinfo {pages} {1539}
  (\bibinfo {year} {2010})}\BibitemShut {NoStop}%
\bibitem [{\citenamefont {Martin}\ and\ \citenamefont
  {Batista}(2008)}]{martin_scalar_2008}%
  \BibitemOpen
  \bibfield  {author} {\bibinfo {author} {\bibfnamefont {I.}~\bibnamefont
  {Martin}}\ and\ \bibinfo {author} {\bibfnamefont {C.~D.}\ \bibnamefont
  {Batista}},\ }\href {\doibase 10.1103/PhysRevLett.101.156402} {\bibfield
  {journal} {\bibinfo  {journal} {Phys. Rev. Lett.}\ }\textbf {\bibinfo
  {volume} {101}},\ \bibinfo {pages} {156402} (\bibinfo {year}
  {2008})}\BibitemShut {NoStop}%
\bibitem [{\citenamefont {Park}\ \emph {et~al.}(2022)\citenamefont {Park},
  \citenamefont {Heinonen},\ and\ \citenamefont
  {Martin}}]{PhysRevMaterials.6.024201}%
  \BibitemOpen
  \bibfield  {author} {\bibinfo {author} {\bibfnamefont {H.}~\bibnamefont
  {Park}}, \bibinfo {author} {\bibfnamefont {O.}~\bibnamefont {Heinonen}}, \
  and\ \bibinfo {author} {\bibfnamefont {I.}~\bibnamefont {Martin}},\ }\href
  {\doibase 10.1103/PhysRevMaterials.6.024201} {\bibfield  {journal} {\bibinfo
  {journal} {Phys. Rev. Mater.}\ }\textbf {\bibinfo {volume} {6}},\ \bibinfo
  {pages} {024201} (\bibinfo {year} {2022})}\BibitemShut {NoStop}%
\bibitem [{\citenamefont {Yang}\ \emph {et~al.}(2022)\citenamefont {Yang},
  \citenamefont {LaBollita}, \citenamefont {Cheng}, \citenamefont {Bhandari},
  \citenamefont {Cochran}, \citenamefont {Yin}, \citenamefont {Hossain},
  \citenamefont {Belopolski}, \citenamefont {Zhang}, \citenamefont {Jiang},
  \citenamefont {Shumiya}, \citenamefont {Multer}, \citenamefont {Liskevich},
  \citenamefont {Usanov}, \citenamefont {Dang}, \citenamefont {Strocov},
  \citenamefont {Davydov}, \citenamefont {Ghimire}, \citenamefont {Botana},\
  and\ \citenamefont {Hasan}}]{PhysRevB.105.L121107}%
  \BibitemOpen
  \bibfield  {author} {\bibinfo {author} {\bibfnamefont {X.~P.}\ \bibnamefont
  {Yang}}, \bibinfo {author} {\bibfnamefont {H.}~\bibnamefont {LaBollita}},
  \bibinfo {author} {\bibfnamefont {Z.-J.}\ \bibnamefont {Cheng}}, \bibinfo
  {author} {\bibfnamefont {H.}~\bibnamefont {Bhandari}}, \bibinfo {author}
  {\bibfnamefont {T.~A.}\ \bibnamefont {Cochran}}, \bibinfo {author}
  {\bibfnamefont {J.-X.}\ \bibnamefont {Yin}}, \bibinfo {author} {\bibfnamefont
  {M.~S.}\ \bibnamefont {Hossain}}, \bibinfo {author} {\bibfnamefont
  {I.}~\bibnamefont {Belopolski}}, \bibinfo {author} {\bibfnamefont
  {Q.}~\bibnamefont {Zhang}}, \bibinfo {author} {\bibfnamefont
  {Y.}~\bibnamefont {Jiang}}, \bibinfo {author} {\bibfnamefont
  {N.}~\bibnamefont {Shumiya}}, \bibinfo {author} {\bibfnamefont
  {D.}~\bibnamefont {Multer}}, \bibinfo {author} {\bibfnamefont
  {M.}~\bibnamefont {Liskevich}}, \bibinfo {author} {\bibfnamefont {D.~A.}\
  \bibnamefont {Usanov}}, \bibinfo {author} {\bibfnamefont {Y.}~\bibnamefont
  {Dang}}, \bibinfo {author} {\bibfnamefont {V.~N.}\ \bibnamefont {Strocov}},
  \bibinfo {author} {\bibfnamefont {A.~V.}\ \bibnamefont {Davydov}}, \bibinfo
  {author} {\bibfnamefont {N.~J.}\ \bibnamefont {Ghimire}}, \bibinfo {author}
  {\bibfnamefont {A.~S.}\ \bibnamefont {Botana}}, \ and\ \bibinfo {author}
  {\bibfnamefont {M.~Z.}\ \bibnamefont {Hasan}},\ }\href {\doibase
  10.1103/PhysRevB.105.L121107} {\bibfield  {journal} {\bibinfo  {journal}
  {Phys. Rev. B}\ }\textbf {\bibinfo {volume} {105}},\ \bibinfo {pages}
  {L121107} (\bibinfo {year} {2022})}\BibitemShut {NoStop}%
\bibitem [{\citenamefont {Pop\ifmmode \check{c}\else
  \v{c}\fi{}evi\ifmmode~\acute{c}\else \'{c}\fi{}}\ \emph
  {et~al.}(2022)\citenamefont {Pop\ifmmode \check{c}\else
  \v{c}\fi{}evi\ifmmode~\acute{c}\else \'{c}\fi{}}, \citenamefont {Utsumi},
  \citenamefont {Bia\l{}o}, \citenamefont {Tabis}, \citenamefont {Gala},
  \citenamefont {Rosmus}, \citenamefont {Kolodziej}, \citenamefont
  {Tomaszewska}, \citenamefont {Garb}, \citenamefont {Berger}, \citenamefont
  {Batisti\ifmmode~\acute{c}\else \'{c}\fi{}}, \citenamefont {Bari\ifmmode
  \check{s}\else \v{s}\fi{}i\ifmmode~\acute{c}\else \'{c}\fi{}}, \citenamefont
  {Forr\'o},\ and\ \citenamefont {Tuti\ifmmode~\check{s}\else
  \v{s}\fi{}}}]{PhysRevB.105.155114}%
  \BibitemOpen
  \bibfield  {author} {\bibinfo {author} {\bibfnamefont {P.}~\bibnamefont
  {Pop\ifmmode \check{c}\else \v{c}\fi{}evi\ifmmode~\acute{c}\else
  \'{c}\fi{}}}, \bibinfo {author} {\bibfnamefont {Y.}~\bibnamefont {Utsumi}},
  \bibinfo {author} {\bibfnamefont {I.}~\bibnamefont {Bia\l{}o}}, \bibinfo
  {author} {\bibfnamefont {W.}~\bibnamefont {Tabis}}, \bibinfo {author}
  {\bibfnamefont {M.~A.}\ \bibnamefont {Gala}}, \bibinfo {author}
  {\bibfnamefont {M.}~\bibnamefont {Rosmus}}, \bibinfo {author} {\bibfnamefont
  {J.~J.}\ \bibnamefont {Kolodziej}}, \bibinfo {author} {\bibfnamefont
  {N.}~\bibnamefont {Tomaszewska}}, \bibinfo {author} {\bibfnamefont
  {M.}~\bibnamefont {Garb}}, \bibinfo {author} {\bibfnamefont {H.}~\bibnamefont
  {Berger}}, \bibinfo {author} {\bibfnamefont {I.}~\bibnamefont
  {Batisti\ifmmode~\acute{c}\else \'{c}\fi{}}}, \bibinfo {author}
  {\bibfnamefont {N.}~\bibnamefont {Bari\ifmmode \check{s}\else
  \v{s}\fi{}i\ifmmode~\acute{c}\else \'{c}\fi{}}}, \bibinfo {author}
  {\bibfnamefont {L.}~\bibnamefont {Forr\'o}}, \ and\ \bibinfo {author}
  {\bibfnamefont {E.}~\bibnamefont {Tuti\ifmmode~\check{s}\else \v{s}\fi{}}},\
  }\href {\doibase 10.1103/PhysRevB.105.155114} {\bibfield  {journal} {\bibinfo
   {journal} {Phys. Rev. B}\ }\textbf {\bibinfo {volume} {105}},\ \bibinfo
  {pages} {155114} (\bibinfo {year} {2022})}\BibitemShut {NoStop}%
\bibitem [{\citenamefont {Tanaka}\ \emph {et~al.}(2022)\citenamefont {Tanaka},
  \citenamefont {Okazaki}, \citenamefont {Kuroda}, \citenamefont {Noguchi},
  \citenamefont {Arai}, \citenamefont {Minami}, \citenamefont {Ideta},
  \citenamefont {Tanaka}, \citenamefont {Lu}, \citenamefont {Hashimoto},
  \citenamefont {Kandyba}, \citenamefont {Cattelan}, \citenamefont {Barinov},
  \citenamefont {Muro}, \citenamefont {Sasagawa},\ and\ \citenamefont
  {Kondo}}]{PhysRevB.105.L121102}%
  \BibitemOpen
  \bibfield  {author} {\bibinfo {author} {\bibfnamefont {H.}~\bibnamefont
  {Tanaka}}, \bibinfo {author} {\bibfnamefont {S.}~\bibnamefont {Okazaki}},
  \bibinfo {author} {\bibfnamefont {K.}~\bibnamefont {Kuroda}}, \bibinfo
  {author} {\bibfnamefont {R.}~\bibnamefont {Noguchi}}, \bibinfo {author}
  {\bibfnamefont {Y.}~\bibnamefont {Arai}}, \bibinfo {author} {\bibfnamefont
  {S.}~\bibnamefont {Minami}}, \bibinfo {author} {\bibfnamefont
  {S.}~\bibnamefont {Ideta}}, \bibinfo {author} {\bibfnamefont
  {K.}~\bibnamefont {Tanaka}}, \bibinfo {author} {\bibfnamefont
  {D.}~\bibnamefont {Lu}}, \bibinfo {author} {\bibfnamefont {M.}~\bibnamefont
  {Hashimoto}}, \bibinfo {author} {\bibfnamefont {V.}~\bibnamefont {Kandyba}},
  \bibinfo {author} {\bibfnamefont {M.}~\bibnamefont {Cattelan}}, \bibinfo
  {author} {\bibfnamefont {A.}~\bibnamefont {Barinov}}, \bibinfo {author}
  {\bibfnamefont {T.}~\bibnamefont {Muro}}, \bibinfo {author} {\bibfnamefont
  {T.}~\bibnamefont {Sasagawa}}, \ and\ \bibinfo {author} {\bibfnamefont
  {T.}~\bibnamefont {Kondo}},\ }\href {\doibase 10.1103/PhysRevB.105.L121102}
  {\bibfield  {journal} {\bibinfo  {journal} {Phys. Rev. B}\ }\textbf {\bibinfo
  {volume} {105}},\ \bibinfo {pages} {L121102} (\bibinfo {year}
  {2022})}\BibitemShut {NoStop}%
\bibitem [{\citenamefont {Parkin}\ \emph {et~al.}(1983)\citenamefont {Parkin},
  \citenamefont {Marseglia},\ and\ \citenamefont
  {Brown}}]{parkin_magnetic_1983}%
  \BibitemOpen
  \bibfield  {author} {\bibinfo {author} {\bibfnamefont {S.~S.~P.}\
  \bibnamefont {Parkin}}, \bibinfo {author} {\bibfnamefont {E.~A.}\
  \bibnamefont {Marseglia}}, \ and\ \bibinfo {author} {\bibfnamefont {P.~J.}\
  \bibnamefont {Brown}},\ }\href {\doibase 10.1088/0022-3719/16/14/016}
  {\bibfield  {journal} {\bibinfo  {journal} {J. Phys. C: Solid State Phys.}\
  }\textbf {\bibinfo {volume} {16}},\ \bibinfo {pages} {2765} (\bibinfo {year}
  {1983})}\BibitemShut {NoStop}%
\bibitem [{\citenamefont {Takagi}\ \emph {et~al.}(2023)\citenamefont {Takagi},
  \citenamefont {Takagi}, \citenamefont {Minami}, \citenamefont {Nomoto},
  \citenamefont {Ohishi}, \citenamefont {Suzuki}, \citenamefont {Yanagi},
  \citenamefont {Hirayama}, \citenamefont {Khanh}, \citenamefont {Karube},
  \citenamefont {Saito}, \citenamefont {Hashizume}, \citenamefont {Kiyanagi},
  \citenamefont {Tokura}, \citenamefont {Arita}, \citenamefont {Nakajima},\
  and\ \citenamefont {Seki}}]{NaturePhysics_Takagi}%
  \BibitemOpen
  \bibfield  {author} {\bibinfo {author} {\bibfnamefont {H.}~\bibnamefont
  {Takagi}}, \bibinfo {author} {\bibfnamefont {R.}~\bibnamefont {Takagi}},
  \bibinfo {author} {\bibfnamefont {S.}~\bibnamefont {Minami}}, \bibinfo
  {author} {\bibfnamefont {T.}~\bibnamefont {Nomoto}}, \bibinfo {author}
  {\bibfnamefont {K.}~\bibnamefont {Ohishi}}, \bibinfo {author} {\bibfnamefont
  {M.-T.}\ \bibnamefont {Suzuki}}, \bibinfo {author} {\bibfnamefont
  {Y.}~\bibnamefont {Yanagi}}, \bibinfo {author} {\bibfnamefont
  {M.}~\bibnamefont {Hirayama}}, \bibinfo {author} {\bibfnamefont {N.~D.}\
  \bibnamefont {Khanh}}, \bibinfo {author} {\bibfnamefont {K.}~\bibnamefont
  {Karube}}, \bibinfo {author} {\bibfnamefont {H.}~\bibnamefont {Saito}},
  \bibinfo {author} {\bibfnamefont {D.}~\bibnamefont {Hashizume}}, \bibinfo
  {author} {\bibfnamefont {R.}~\bibnamefont {Kiyanagi}}, \bibinfo {author}
  {\bibfnamefont {Y.}~\bibnamefont {Tokura}}, \bibinfo {author} {\bibfnamefont
  {R.}~\bibnamefont {Arita}}, \bibinfo {author} {\bibfnamefont
  {T.}~\bibnamefont {Nakajima}}, \ and\ \bibinfo {author} {\bibfnamefont
  {S.}~\bibnamefont {Seki}},\ }\href {\doibase 10.1038/s41567-023-02017-3}
  {\bibfield  {journal} {\bibinfo  {journal} {Nature Physics}\ }\textbf
  {\bibinfo {volume} {19}},\ \bibinfo {pages} {961} (\bibinfo {year}
  {2023})}\BibitemShut {NoStop}%
\bibitem [{\citenamefont {Batista}\ \emph {et~al.}(2016)\citenamefont
  {Batista}, \citenamefont {Lin}, \citenamefont {Hayami},\ and\ \citenamefont
  {Kamiya}}]{batista2016frustration}%
  \BibitemOpen
  \bibfield  {author} {\bibinfo {author} {\bibfnamefont {C.~D.}\ \bibnamefont
  {Batista}}, \bibinfo {author} {\bibfnamefont {S.-Z.}\ \bibnamefont {Lin}},
  \bibinfo {author} {\bibfnamefont {S.}~\bibnamefont {Hayami}}, \ and\ \bibinfo
  {author} {\bibfnamefont {Y.}~\bibnamefont {Kamiya}},\ }\href@noop {}
  {\bibfield  {journal} {\bibinfo  {journal} {Reports on Progress in Physics}\
  }\textbf {\bibinfo {volume} {79}},\ \bibinfo {pages} {084504} (\bibinfo
  {year} {2016})}\BibitemShut {NoStop}%
\bibitem [{\citenamefont {Kresse}\ and\ \citenamefont
  {Furthm\"uller}(1996)}]{vasp1}%
  \BibitemOpen
  \bibfield  {author} {\bibinfo {author} {\bibfnamefont {G.}~\bibnamefont
  {Kresse}}\ and\ \bibinfo {author} {\bibfnamefont {J.}~\bibnamefont
  {Furthm\"uller}},\ }\href {\doibase 10.1103/PhysRevB.54.11169} {\bibfield
  {journal} {\bibinfo  {journal} {Phys. Rev. B}\ }\textbf {\bibinfo {volume}
  {54}},\ \bibinfo {pages} {11169} (\bibinfo {year} {1996})}\BibitemShut
  {NoStop}%
\bibitem [{\citenamefont {Kresse}\ and\ \citenamefont {Joubert}(1999)}]{vasp2}%
  \BibitemOpen
  \bibfield  {author} {\bibinfo {author} {\bibfnamefont {G.}~\bibnamefont
  {Kresse}}\ and\ \bibinfo {author} {\bibfnamefont {D.}~\bibnamefont
  {Joubert}},\ }\href {\doibase 10.1103/PhysRevB.59.1758} {\bibfield  {journal}
  {\bibinfo  {journal} {Phys. Rev. B}\ }\textbf {\bibinfo {volume} {59}},\
  \bibinfo {pages} {1758} (\bibinfo {year} {1999})}\BibitemShut {NoStop}%
\bibitem [{\citenamefont {Marzari}\ and\ \citenamefont
  {Vanderbilt}(1997)}]{marzari1997maximally}%
  \BibitemOpen
  \bibfield  {author} {\bibinfo {author} {\bibfnamefont {N.}~\bibnamefont
  {Marzari}}\ and\ \bibinfo {author} {\bibfnamefont {D.}~\bibnamefont
  {Vanderbilt}},\ }\href@noop {} {\bibfield  {journal} {\bibinfo  {journal}
  {Physical review B}\ }\textbf {\bibinfo {volume} {56}},\ \bibinfo {pages}
  {12847} (\bibinfo {year} {1997})}\BibitemShut {NoStop}%
\bibitem [{\citenamefont {Singh}\ \emph {et~al.}(2021)\citenamefont {Singh},
  \citenamefont {Herath}, \citenamefont {Wah}, \citenamefont {Liao},
  \citenamefont {Romero},\ and\ \citenamefont {Park}}]{singh2021dmftwdft}%
  \BibitemOpen
  \bibfield  {author} {\bibinfo {author} {\bibfnamefont {V.}~\bibnamefont
  {Singh}}, \bibinfo {author} {\bibfnamefont {U.}~\bibnamefont {Herath}},
  \bibinfo {author} {\bibfnamefont {B.}~\bibnamefont {Wah}}, \bibinfo {author}
  {\bibfnamefont {X.}~\bibnamefont {Liao}}, \bibinfo {author} {\bibfnamefont
  {A.~H.}\ \bibnamefont {Romero}}, \ and\ \bibinfo {author} {\bibfnamefont
  {H.}~\bibnamefont {Park}},\ }\href@noop {} {\bibfield  {journal} {\bibinfo
  {journal} {Computer Physics Communications}\ }\textbf {\bibinfo {volume}
  {261}},\ \bibinfo {pages} {107778} (\bibinfo {year} {2021})}\BibitemShut
  {NoStop}%
\bibitem [{\citenamefont {Haule}(2007)}]{ctqmc1}%
  \BibitemOpen
  \bibfield  {author} {\bibinfo {author} {\bibfnamefont {K.}~\bibnamefont
  {Haule}},\ }\href {\doibase 10.1103/PhysRevB.75.155113} {\bibfield  {journal}
  {\bibinfo  {journal} {Phys. Rev. B}\ }\textbf {\bibinfo {volume} {75}},\
  \bibinfo {pages} {155113} (\bibinfo {year} {2007})}\BibitemShut {NoStop}%
\bibitem [{\citenamefont {Heil}\ \emph {et~al.}(2014)\citenamefont {Heil},
  \citenamefont {Sormann}, \citenamefont {Boeri}, \citenamefont {Aichhorn},\
  and\ \citenamefont {von~der Linden}}]{PhysRevB.90.115143}%
  \BibitemOpen
  \bibfield  {author} {\bibinfo {author} {\bibfnamefont {C.}~\bibnamefont
  {Heil}}, \bibinfo {author} {\bibfnamefont {H.}~\bibnamefont {Sormann}},
  \bibinfo {author} {\bibfnamefont {L.}~\bibnamefont {Boeri}}, \bibinfo
  {author} {\bibfnamefont {M.}~\bibnamefont {Aichhorn}}, \ and\ \bibinfo
  {author} {\bibfnamefont {W.}~\bibnamefont {von~der Linden}},\ }\href
  {\doibase 10.1103/PhysRevB.90.115143} {\bibfield  {journal} {\bibinfo
  {journal} {Phys. Rev. B}\ }\textbf {\bibinfo {volume} {90}},\ \bibinfo
  {pages} {115143} (\bibinfo {year} {2014})}\BibitemShut {NoStop}%
\bibitem [{\citenamefont {Park}\ \emph {et~al.}(2011)\citenamefont {Park},
  \citenamefont {Haule},\ and\ \citenamefont
  {Kotliar}}]{PhysRevLett.107.137007}%
  \BibitemOpen
  \bibfield  {author} {\bibinfo {author} {\bibfnamefont {H.}~\bibnamefont
  {Park}}, \bibinfo {author} {\bibfnamefont {K.}~\bibnamefont {Haule}}, \ and\
  \bibinfo {author} {\bibfnamefont {G.}~\bibnamefont {Kotliar}},\ }\href
  {\doibase 10.1103/PhysRevLett.107.137007} {\bibfield  {journal} {\bibinfo
  {journal} {Phys. Rev. Lett.}\ }\textbf {\bibinfo {volume} {107}},\ \bibinfo
  {pages} {137007} (\bibinfo {year} {2011})}\BibitemShut {NoStop}%
\bibitem [{\citenamefont {Goremychkin}\ \emph {et~al.}(2018)\citenamefont
  {Goremychkin}, \citenamefont {Park}, \citenamefont {Osborn}, \citenamefont
  {Rosenkranz}, \citenamefont {Castellan}, \citenamefont {Fanelli},
  \citenamefont {Christianson}, \citenamefont {Stone}, \citenamefont {Bauer},
  \citenamefont {McClellan}, \citenamefont {Byler},\ and\ \citenamefont
  {Lawrence}}]{doi:10.1126/science.aan0593}%
  \BibitemOpen
  \bibfield  {author} {\bibinfo {author} {\bibfnamefont {E.~A.}\ \bibnamefont
  {Goremychkin}}, \bibinfo {author} {\bibfnamefont {H.}~\bibnamefont {Park}},
  \bibinfo {author} {\bibfnamefont {R.}~\bibnamefont {Osborn}}, \bibinfo
  {author} {\bibfnamefont {S.}~\bibnamefont {Rosenkranz}}, \bibinfo {author}
  {\bibfnamefont {J.-P.}\ \bibnamefont {Castellan}}, \bibinfo {author}
  {\bibfnamefont {V.~R.}\ \bibnamefont {Fanelli}}, \bibinfo {author}
  {\bibfnamefont {A.~D.}\ \bibnamefont {Christianson}}, \bibinfo {author}
  {\bibfnamefont {M.~B.}\ \bibnamefont {Stone}}, \bibinfo {author}
  {\bibfnamefont {E.~D.}\ \bibnamefont {Bauer}}, \bibinfo {author}
  {\bibfnamefont {K.~J.}\ \bibnamefont {McClellan}}, \bibinfo {author}
  {\bibfnamefont {D.~D.}\ \bibnamefont {Byler}}, \ and\ \bibinfo {author}
  {\bibfnamefont {J.~M.}\ \bibnamefont {Lawrence}},\ }\href {\doibase
  10.1126/science.aan0593} {\bibfield  {journal} {\bibinfo  {journal}
  {Science}\ }\textbf {\bibinfo {volume} {359}},\ \bibinfo {pages} {186}
  (\bibinfo {year} {2018})}\BibitemShut {NoStop}%
\bibitem [{\citenamefont {Johannes}\ \emph {et~al.}(2006)\citenamefont
  {Johannes}, \citenamefont {Mazin},\ and\ \citenamefont
  {Howells}}]{PhysRevB.73.205102}%
  \BibitemOpen
  \bibfield  {author} {\bibinfo {author} {\bibfnamefont {M.~D.}\ \bibnamefont
  {Johannes}}, \bibinfo {author} {\bibfnamefont {I.~I.}\ \bibnamefont {Mazin}},
  \ and\ \bibinfo {author} {\bibfnamefont {C.~A.}\ \bibnamefont {Howells}},\
  }\href {\doibase 10.1103/PhysRevB.73.205102} {\bibfield  {journal} {\bibinfo
  {journal} {Phys. Rev. B}\ }\textbf {\bibinfo {volume} {73}},\ \bibinfo
  {pages} {205102} (\bibinfo {year} {2006})}\BibitemShut {NoStop}%
\bibitem [{\citenamefont {Lane}\ \emph {et~al.}(2023)\citenamefont {Lane},
  \citenamefont {Zhang}, \citenamefont {Barbiellini}, \citenamefont
  {Markiewicz}, \citenamefont {Bansil}, \citenamefont {Sun},\ and\
  \citenamefont {Zhu}}]{Chris_susceptibility}%
  \BibitemOpen
  \bibfield  {author} {\bibinfo {author} {\bibfnamefont {C.}~\bibnamefont
  {Lane}}, \bibinfo {author} {\bibfnamefont {R.}~\bibnamefont {Zhang}},
  \bibinfo {author} {\bibfnamefont {B.}~\bibnamefont {Barbiellini}}, \bibinfo
  {author} {\bibfnamefont {R.~S.}\ \bibnamefont {Markiewicz}}, \bibinfo
  {author} {\bibfnamefont {A.}~\bibnamefont {Bansil}}, \bibinfo {author}
  {\bibfnamefont {J.}~\bibnamefont {Sun}}, \ and\ \bibinfo {author}
  {\bibfnamefont {J.-X.}\ \bibnamefont {Zhu}},\ }\href@noop {} {\bibfield
  {journal} {\bibinfo  {journal} {Communications Physics}\ }\textbf {\bibinfo
  {volume} {6}},\ \bibinfo {pages} {90} (\bibinfo {year} {2023})}\BibitemShut
  {NoStop}%
\bibitem [{\citenamefont {Takagi}\ \emph {et~al.}(2018)\citenamefont {Takagi},
  \citenamefont {White}, \citenamefont {Hayami}, \citenamefont {Arita},
  \citenamefont {Honecker}, \citenamefont {Rønnow}, \citenamefont {Tokura},\
  and\ \citenamefont {Seki}}]{doi:10.1126/sciadv.aau3402}%
  \BibitemOpen
  \bibfield  {author} {\bibinfo {author} {\bibfnamefont {R.}~\bibnamefont
  {Takagi}}, \bibinfo {author} {\bibfnamefont {J.~S.}\ \bibnamefont {White}},
  \bibinfo {author} {\bibfnamefont {S.}~\bibnamefont {Hayami}}, \bibinfo
  {author} {\bibfnamefont {R.}~\bibnamefont {Arita}}, \bibinfo {author}
  {\bibfnamefont {D.}~\bibnamefont {Honecker}}, \bibinfo {author}
  {\bibfnamefont {H.~M.}\ \bibnamefont {Rønnow}}, \bibinfo {author}
  {\bibfnamefont {Y.}~\bibnamefont {Tokura}}, \ and\ \bibinfo {author}
  {\bibfnamefont {S.}~\bibnamefont {Seki}},\ }\href {\doibase
  10.1126/sciadv.aau3402} {\bibfield  {journal} {\bibinfo  {journal} {Science
  Advances}\ }\textbf {\bibinfo {volume} {4}},\ \bibinfo {pages} {eaau3402}
  (\bibinfo {year} {2018})}\BibitemShut {NoStop}%
\bibitem [{\citenamefont {Bisogni}\ \emph {et~al.}(2016)\citenamefont
  {Bisogni}, \citenamefont {Catalano}, \citenamefont {Green}, \citenamefont
  {Gibert}, \citenamefont {Scherwitzl}, \citenamefont {Huang}, \citenamefont
  {Strocov}, \citenamefont {Zubko}, \citenamefont {Balandeh}, \citenamefont
  {Triscone}, \citenamefont {Sawatzky},\ and\ \citenamefont
  {Schmitt}}]{nickelate}%
  \BibitemOpen
  \bibfield  {author} {\bibinfo {author} {\bibfnamefont {V.}~\bibnamefont
  {Bisogni}}, \bibinfo {author} {\bibfnamefont {S.}~\bibnamefont {Catalano}},
  \bibinfo {author} {\bibfnamefont {R.~J.}\ \bibnamefont {Green}}, \bibinfo
  {author} {\bibfnamefont {M.}~\bibnamefont {Gibert}}, \bibinfo {author}
  {\bibfnamefont {R.}~\bibnamefont {Scherwitzl}}, \bibinfo {author}
  {\bibfnamefont {Y.}~\bibnamefont {Huang}}, \bibinfo {author} {\bibfnamefont
  {V.~N.}\ \bibnamefont {Strocov}}, \bibinfo {author} {\bibfnamefont
  {P.}~\bibnamefont {Zubko}}, \bibinfo {author} {\bibfnamefont
  {S.}~\bibnamefont {Balandeh}}, \bibinfo {author} {\bibfnamefont {J.-M.}\
  \bibnamefont {Triscone}}, \bibinfo {author} {\bibfnamefont {G.}~\bibnamefont
  {Sawatzky}}, \ and\ \bibinfo {author} {\bibfnamefont {T.}~\bibnamefont
  {Schmitt}},\ }\href@noop {} {\bibfield  {journal} {\bibinfo  {journal}
  {Nature Communications}\ }\textbf {\bibinfo {volume} {7}},\ \bibinfo {pages}
  {13017} (\bibinfo {year} {2016})}\BibitemShut {NoStop}%
\bibitem [{\citenamefont {Anzenhofer}\ \emph {et~al.}(1970)\citenamefont
  {Anzenhofer}, \citenamefont {van~den Berg}, \citenamefont {Cossee},\ and\
  \citenamefont {Helle}}]{anzenhofer}%
  \BibitemOpen
  \bibfield  {author} {\bibinfo {author} {\bibfnamefont {K.}~\bibnamefont
  {Anzenhofer}}, \bibinfo {author} {\bibfnamefont {J.}~\bibnamefont {van~den
  Berg}}, \bibinfo {author} {\bibfnamefont {P.}~\bibnamefont {Cossee}}, \ and\
  \bibinfo {author} {\bibfnamefont {J.}~\bibnamefont {Helle}},\ }\href@noop {}
  {\bibfield  {journal} {\bibinfo  {journal} {J. Phys. Chem. Solids}\ }\textbf
  {\bibinfo {volume} {31}},\ \bibinfo {pages} {1057} (\bibinfo {year}
  {1970})}\BibitemShut {NoStop}%
\bibitem [{\citenamefont {Ku}\ \emph {et~al.}(2010)\citenamefont {Ku},
  \citenamefont {Berlijn},\ and\ \citenamefont {Lee}}]{PhysRevLett.104.216401}%
  \BibitemOpen
  \bibfield  {author} {\bibinfo {author} {\bibfnamefont {W.}~\bibnamefont
  {Ku}}, \bibinfo {author} {\bibfnamefont {T.}~\bibnamefont {Berlijn}}, \ and\
  \bibinfo {author} {\bibfnamefont {C.-C.}\ \bibnamefont {Lee}},\ }\href
  {\doibase 10.1103/PhysRevLett.104.216401} {\bibfield  {journal} {\bibinfo
  {journal} {Phys. Rev. Lett.}\ }\textbf {\bibinfo {volume} {104}},\ \bibinfo
  {pages} {216401} (\bibinfo {year} {2010})}\BibitemShut {NoStop}%
\end{thebibliography}%

\section*{Appendix: The derivation and effect of the Form factor}

The Form factor $F(\mathbf{q})$ in Eq.\:\ref{eq:chiqw} can be given by
\begin{eqnarray}
F_{nn'\bar{n}\bar{n}'}(\mathbf{q}) &=& \int d\mathbf{r} \int d\mathbf{r'} e^{i\mathbf{q}(\mathbf{r'}-\mathbf{r})}
\phi^*_{n}(\mathbf{r'})\phi_{n'}(\mathbf{r'})
\phi^*_{\bar{n}}(\mathbf{r})\phi_{\bar{n}'}(\mathbf{r}) \nonumber\\
&=&\langle\phi^{\mathbf{k}}_{n_{\alpha}}|e^{i\mathbf{q}\cdot\hat{\mathbf{r}}'}|\phi^{\mathbf{k'}}_{n_{\alpha'}}\rangle \langle\phi^{\mathbf{\bar{k}}}_{n_{\bar{\alpha}}}|e^{-i\mathbf{q}\cdot\hat{\mathbf{r}}}|\phi^{\bar{\mathbf{k}'}}_{n_{\bar{\alpha}'}}\rangle,
\label{eq:form}
\end{eqnarray}
where the $e^{i\mathbf{q}\cdot\hat{\mathbf{r}}}$ operator can be typically simplified using the dipole approximation or the constant matrix element approximation ($e^{i\mathbf{q}\cdot\hat{\mathbf{r}}}\simeq 1$). 

For a localized orbital $\phi_n(\mathbf{r})$, such as the Wannier function, the expectation value of the $e^{i\mathbf{q}\cdot\hat{\mathbf{r}}}$ factor can be performed using the discrete sum of the center of orbitals.
\begin{eqnarray}
F_{nn'\bar{n}\bar{n}'}(\mathbf{q}) &\simeq& \langle\phi^{\mathbf{k}}_{n_{\alpha}}|e^{i\mathbf{q}\cdot\hat{\mathbf{R}}'}|\phi^{\mathbf{k'}}_{n_{\alpha'}}\rangle \langle\phi^{\mathbf{\bar{k}}}_{n_{\bar{\alpha}}}|e^{-i\mathbf{q}\cdot\hat{\mathbf{R}}}|\phi^{\bar{\mathbf{k}'}}_{n_{\bar{\alpha}'}}\rangle \nonumber \\
&=&\langle\phi^{\mathbf{k+q}}_{n_{\alpha}}|\phi^{\mathbf{k'}}_{n_{\alpha'}}\rangle \langle\psi^{\bar{\mathbf{k}}}_{n_{\bar{\alpha}}}|\psi^{\bar{\mathbf{k}}'+\mathbf{q}}_{n_{\bar{\alpha}'}}\rangle.
\label{eq:form2}
\end{eqnarray}
Since the Wannier function $\phi^{\mathbf{k}}_{n_\alpha}$ is complete and orthonomalized, the Form factor can be generally given by the form of delta functions obeying the momentum and orbital conservation:
\begin{eqnarray}
F_{nn'\bar{n}\bar{n}'}(\mathbf{q})&=& \delta_{\mathbf{k}+\mathbf{q},\mathbf{k'}} \cdot 
\delta_{\bar{\mathbf{k}}'+\mathbf{q},\bar{\mathbf{k}}} \cdot \delta_{n_{\alpha},n_{\alpha'}} \cdot \delta_{n_{\bar{\alpha}},n_{\bar{\alpha}'}}.
\label{eq:delta_form2}
\end{eqnarray}


Here, one should note that the orbital index $n_\alpha$ used in computing the magnetic susceptibility runs over the correlated orbitals with finite magnetic moments, while the generated Wannier functions in the primitive unit cell contain both correlated and non-correlated orbitals.
For example, Co$_{1/3}$NbS$_2$ structure contains two Co ions per primitive unit cell, with the magnetic moments primarily residing on $d$ orbitals of Co ions, while the band structure calculation can generate both the Co $d$ and the Nb $d_{z^2}$ Wannier orbitals as they are strongly hybridized. 
Once the Form factor considers only Co ions, one can define a new momentum vector $\tilde{\mathbf{k}}$ to describe the Wannier function $\tilde{\phi}$ for Co ions only, which will form effectively a single triangular lattice with the new periodicity:
\begin{equation}
\tilde{\phi}^{\tilde{\mathbf{k}}}_{\alpha}(\mathbf{r})=\frac{1}{\sqrt{N N_{\tau}}}\sum_{\mathbf{R},\vec{\tau}_{\alpha}} e^{-i\tilde{\mathbf{k}}\cdot(\mathbf{R}+\vec{\tau}_{\alpha})}\tilde{\phi}^{\tau_{\alpha}}_{\alpha}(\mathbf{r}-\mathbf{R}-\vec{\tau}_{\alpha}).
\end{equation}
Here, the $\tilde{\phi}$ is obtained from the Fourier transform using explicitly the position $\tau_{\alpha}$ of the correlated orbitals. 

To evaluate the matrix element in Eq.\:\ref{eq:form2}, we expand it using the new Wannier orbital $\tilde{\phi}$:
\begin{eqnarray}
\langle\phi^{\mathbf{k}}_{n_\alpha}|e^{i\mathbf{q}\cdot\hat{\mathbf{R}}'}|\phi^{\mathbf{k'}}_{n_{\alpha'}}\rangle &=&  \sum_{\mathbf{\tilde{k}},\tilde{\mathbf{k}}'} \sum_{
\tilde{\alpha},\tilde{\alpha}'}\langle\phi^{\mathbf{k}}_{n_\alpha}|\tilde{\phi}^{\tilde{\mathbf{k}}}_{\tilde{\alpha}}\rangle\langle\tilde{\phi}^{\tilde{\mathbf{k}}}_{\tilde{\alpha}}|e^{i\mathbf{q}\cdot\mathbf{\hat{R}'}}|\tilde{\phi}^{\tilde{\mathbf{k}}'}_{\tilde{\alpha}'}\rangle \nonumber\\
&& \cdot\langle\tilde{\phi}^{\tilde{\mathbf{k'}}}_{\tilde{\alpha}'}| \phi^{\mathbf{k'}}_{n_{\alpha'}}\rangle \nonumber\\
&=& \frac{1}{N_{\tau}}\sum_{\mathbf{\tilde{k}},\mathbf{\tilde{k}}'} \sum_{\vec{\tau}_\alpha,\vec{\tau}_{\alpha'}} e^{i(\tilde{\mathbf{k}}'\cdot\vec{\tau}_{\alpha'}-\tilde{\mathbf{k}}\cdot\vec{\tau}_{\alpha})}\cdot
\delta_{\alpha \alpha'}\nonumber\\
&& \cdot\delta_{\{\mathbf{k}\},\tilde{\mathbf{k}}}
\cdot \delta_{\{\mathbf{k}'\},\tilde{\mathbf{k}}'}
\cdot \delta_{\tilde{\mathbf{k}}+\{\mathbf{q}\}, \tilde{\mathbf{k}}'},
\label{eq:form3}
\end{eqnarray}
where 
the momentum $\tilde{\mathbf{k}}$ is defined in an extended BZ obtained for a single Co ion in triangular lattice, therefore the overlap between $\phi^{\mathbf{k}}_{n_\alpha}$ and $\tilde{\phi}^{\tilde{\mathbf{k}}}_{\tilde{\alpha}}$ can produce the modulation factor obtained from the Co ion positions due to the effective band unfolding effect~\cite{PhysRevLett.104.216401}. 

To obtain Eq.\:\ref{eq:form3}, we evaluate the overlap between these orbitals as follows:
\begin{eqnarray}
\langle\phi_{n_\alpha}^{\mathbf{k}}|\tilde{\phi}_{\tilde{\alpha}}^{\tilde{\mathbf{k}}}\rangle &=& \frac{1}{N\sqrt{N_{\tau}}}\sum_{\mathbf{R},\vec{\tau}_{\tilde{\alpha}}} e^{i(\mathbf{k}\cdot\mathbf{R}-\tilde{\mathbf{k}}\cdot(\mathbf{R}+\vec{\tau}_{\tilde{\alpha}}))}\langle\phi_{\alpha}^{\tau_{\alpha}}|\tilde{\phi}_{\tilde{\alpha}}^{\tau_{\tilde{\alpha}}}\rangle \nonumber\\
&=& \frac{1}{\sqrt{N_{\tau}}}\sum_{\vec{\tau}_{\tilde{\alpha}}}e^{-i\tilde{\mathbf{k}}\vec{\tau}_{\tilde{\alpha}}}
\cdot\delta_{\{\mathbf{k}\},\tilde{\mathbf{k}}}
\cdot\delta_{\alpha, \tilde{\alpha}}
\end{eqnarray}
Similarly, 
\begin{eqnarray}
\langle\tilde{\phi}_{\tilde{\alpha}}^{\tilde{\mathbf{k}}}|e^{i\mathbf{q}\cdot\mathbf{\hat{R}'}}|\tilde{\phi}_{\tilde{\alpha}'}^{\tilde{\mathbf{k}}'}\rangle &=& \frac{1}{NN_{\tau}}\sum_{\mathbf{R'},\vec{\tau}_{\tilde{\alpha}}} e^{i(\tilde{\mathbf{k}}-\tilde{\mathbf{k}}'+\mathbf{q})\cdot(\mathbf{R'}+\vec{\tau}_{\tilde{\alpha}})} 
\langle\tilde{\phi}_{\tilde{\alpha}}^{\tau_{\tilde{\alpha}}}|\tilde{\phi}_{\tilde{\alpha}'}^{\tau_{\tilde{\alpha}'}}\rangle\nonumber\\
&=& 
\delta_{\tilde{\mathbf{k}}+\{\mathbf{q}\}, \tilde{\mathbf{k}}'} \cdot
\delta_{\tilde{\alpha}, \tilde{\alpha}'}
\end{eqnarray}

\begin{figure*}[!ht]
\includegraphics[width=0.47\linewidth]{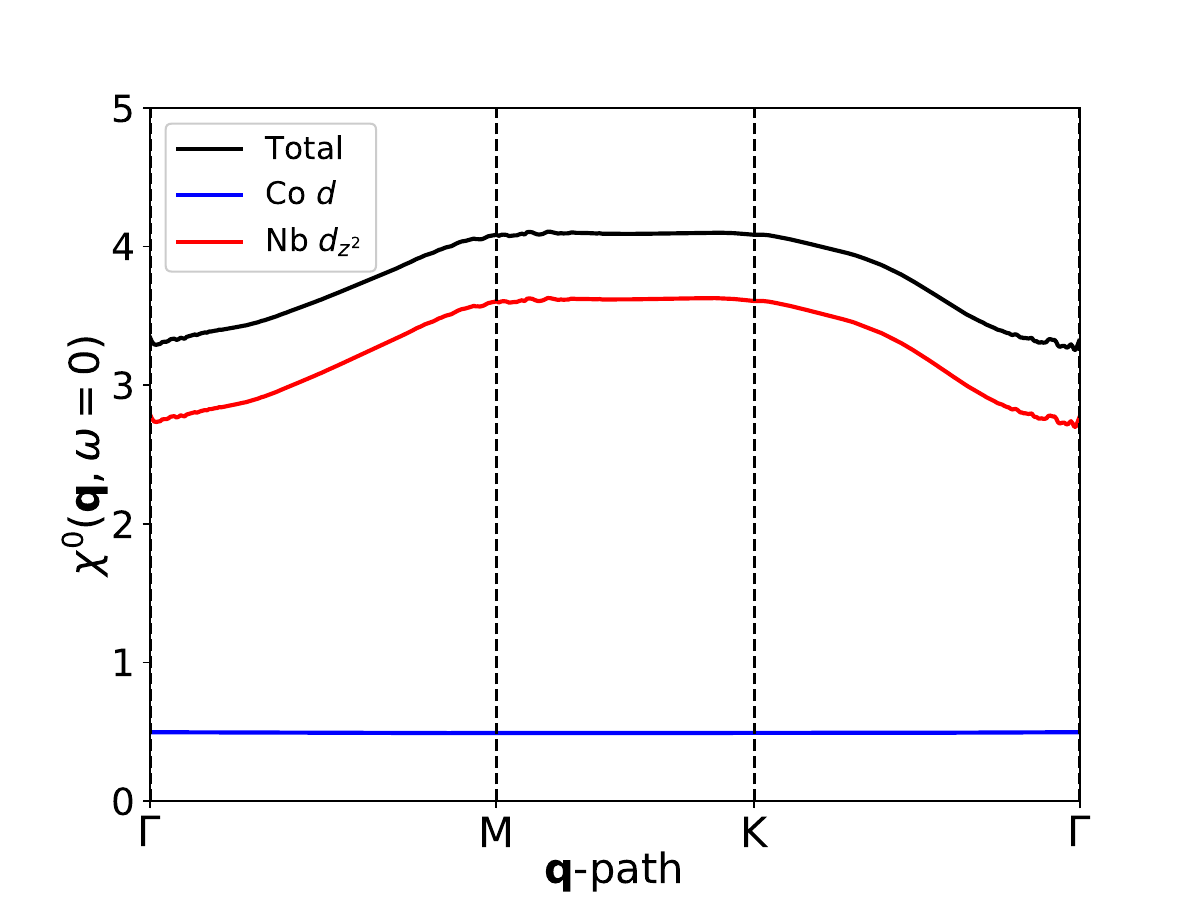}
\includegraphics[width=0.47\linewidth]{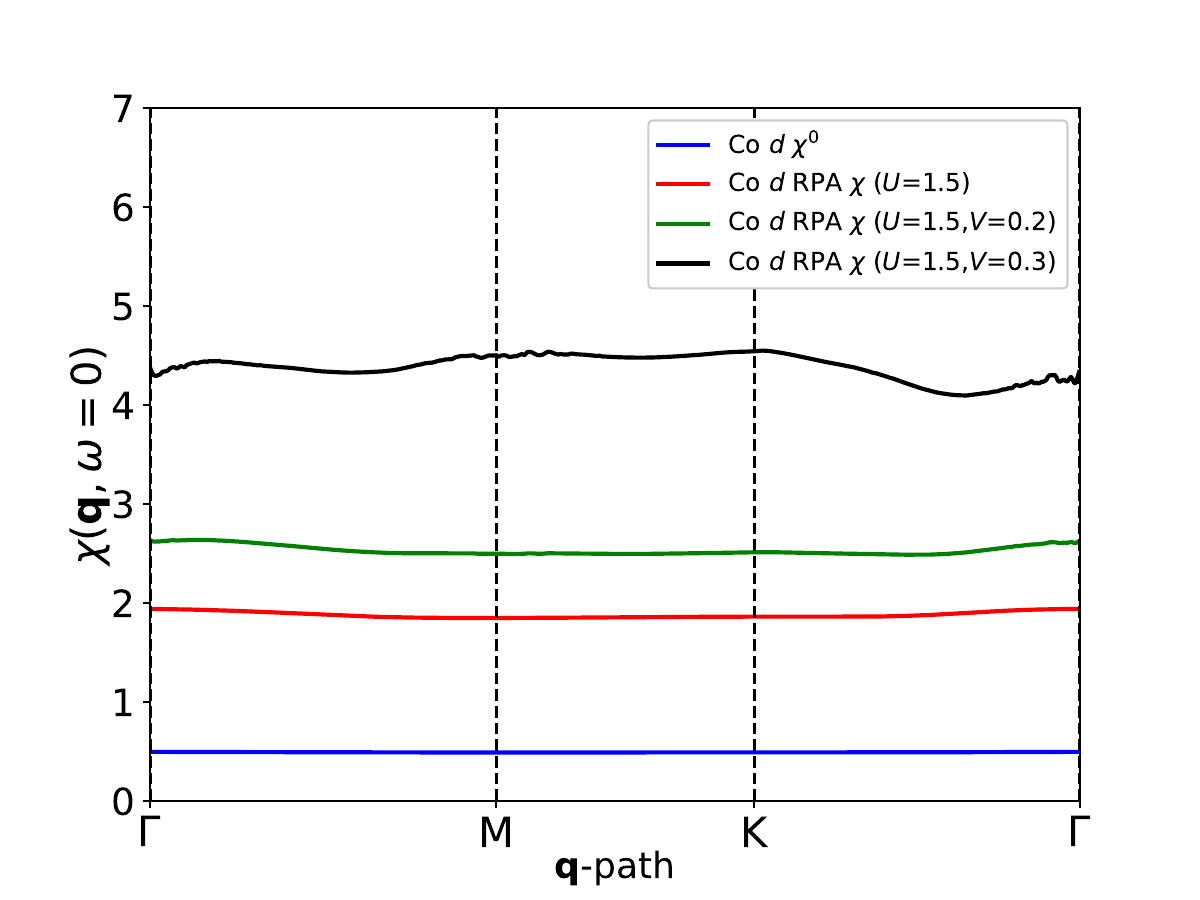}
\caption{The magnetic susceptibility for Co$_{1/3}$NbS$_2$ at the doping $\delta$=2.0, the polarizability $\chi^0$ (left panel) and the RPA $\chi$ (right panel), computed using the form factor in Eq.\:\ref{eq:delta_form2}.
}
\label{fig:Co_chi_no_matrix}
\end{figure*}

\begin{figure*}[!ht]
\includegraphics[width=0.47\linewidth]{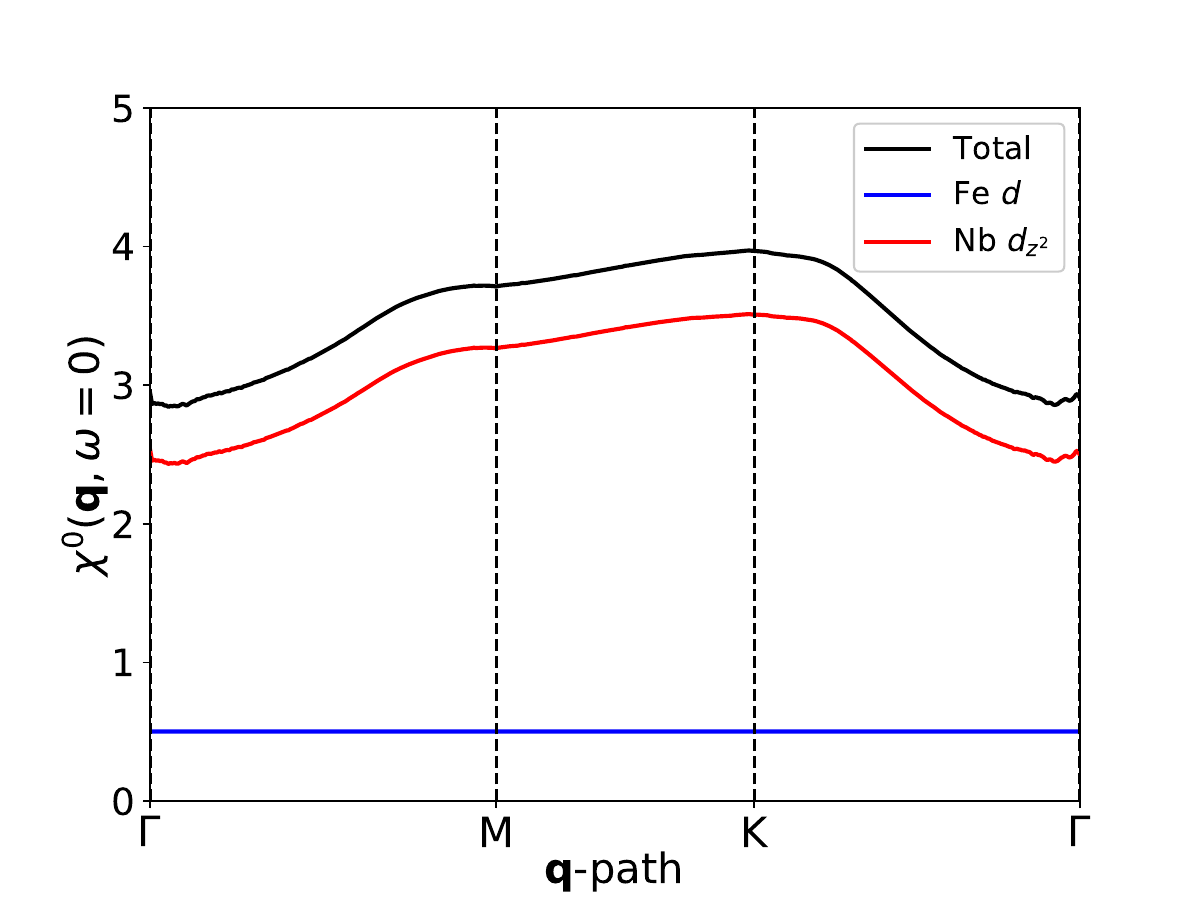}
\includegraphics[width=0.47\linewidth]{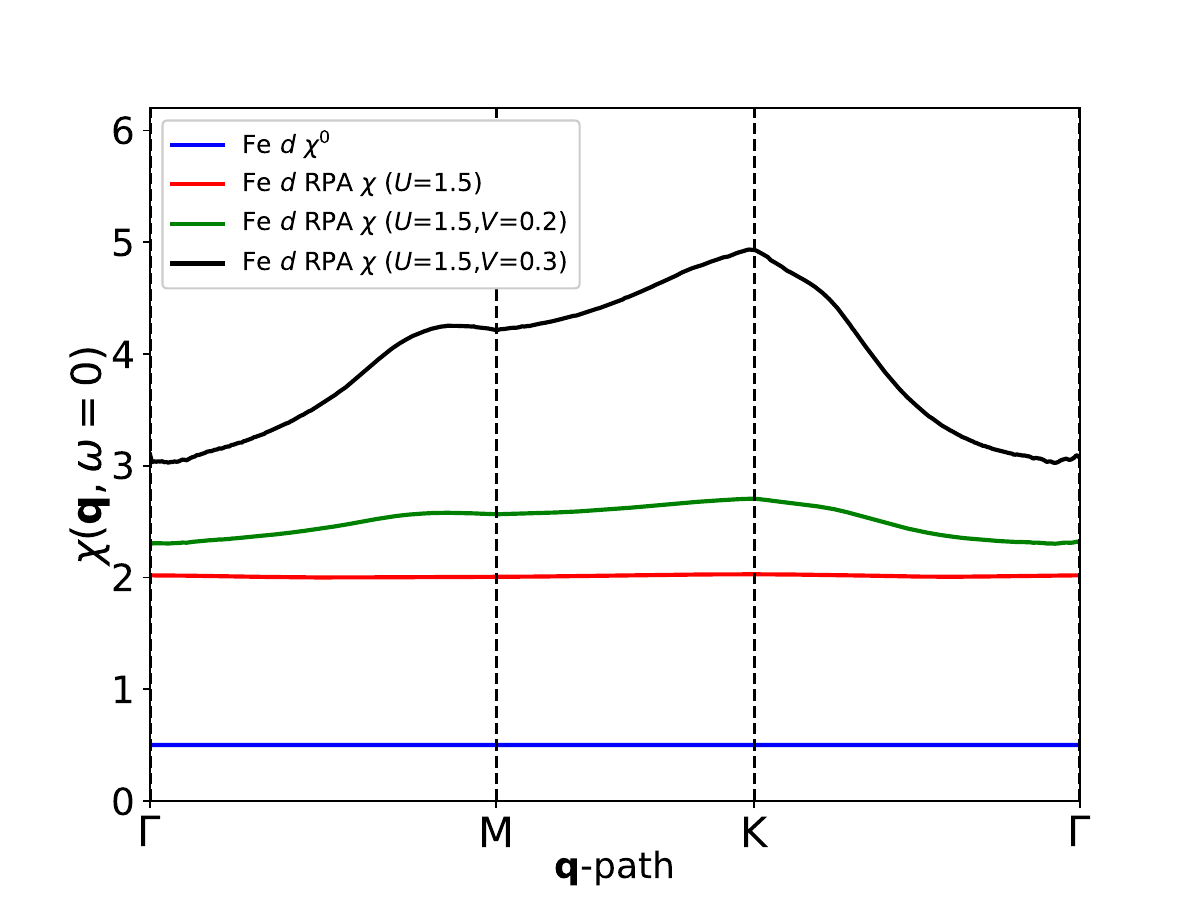}
\caption{The magnetic susceptibility for Fe$_{1/3}$NbS$_2$ at the doping $\delta$=2.0, the polarizability $\chi^0$ (left panel) and the RPA $\chi$ (right panel), computed using the form factor in Eq.\:\ref{eq:delta_form2}.
}
\label{fig:Fe_chi_no_matrix}
\end{figure*}

\begin{figure*}[!ht]
\includegraphics[width=0.47\linewidth]{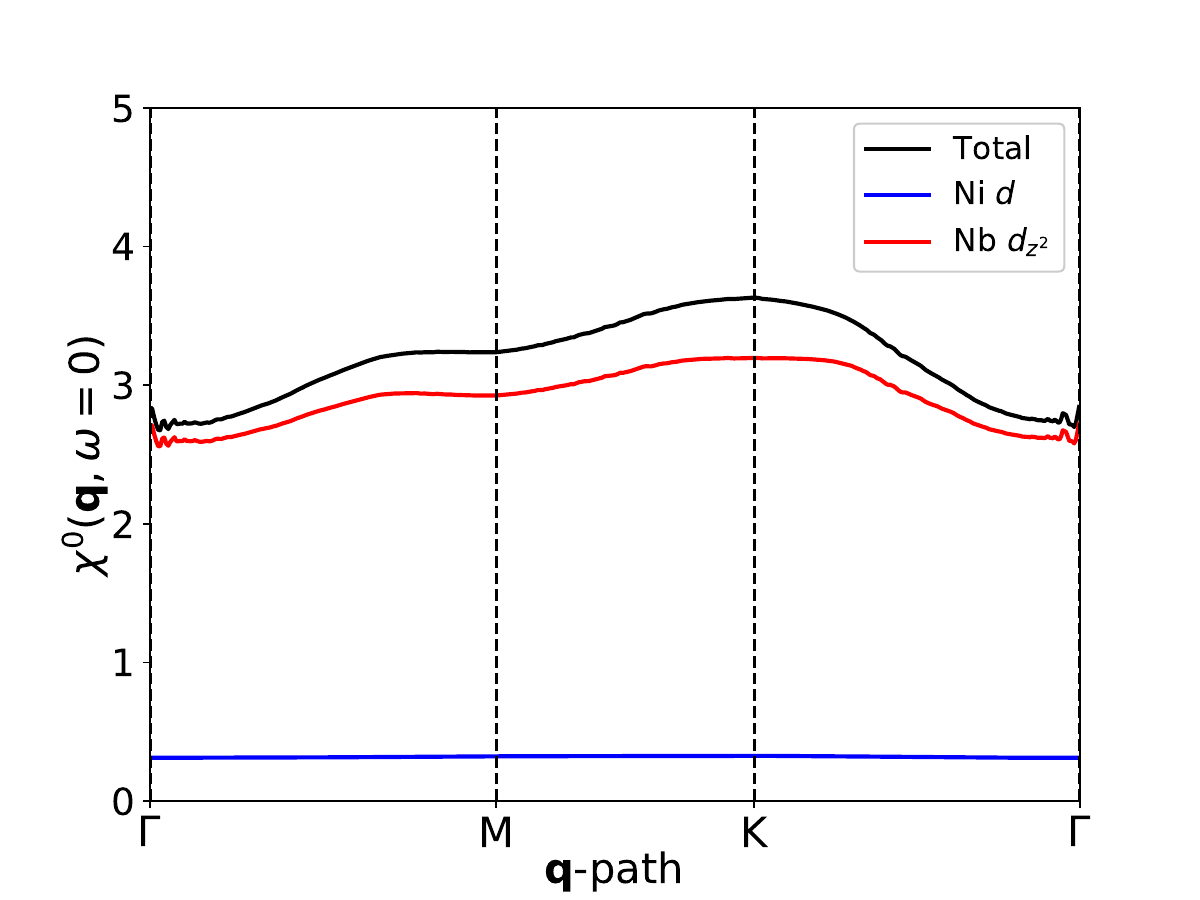}
\includegraphics[width=0.47\linewidth]{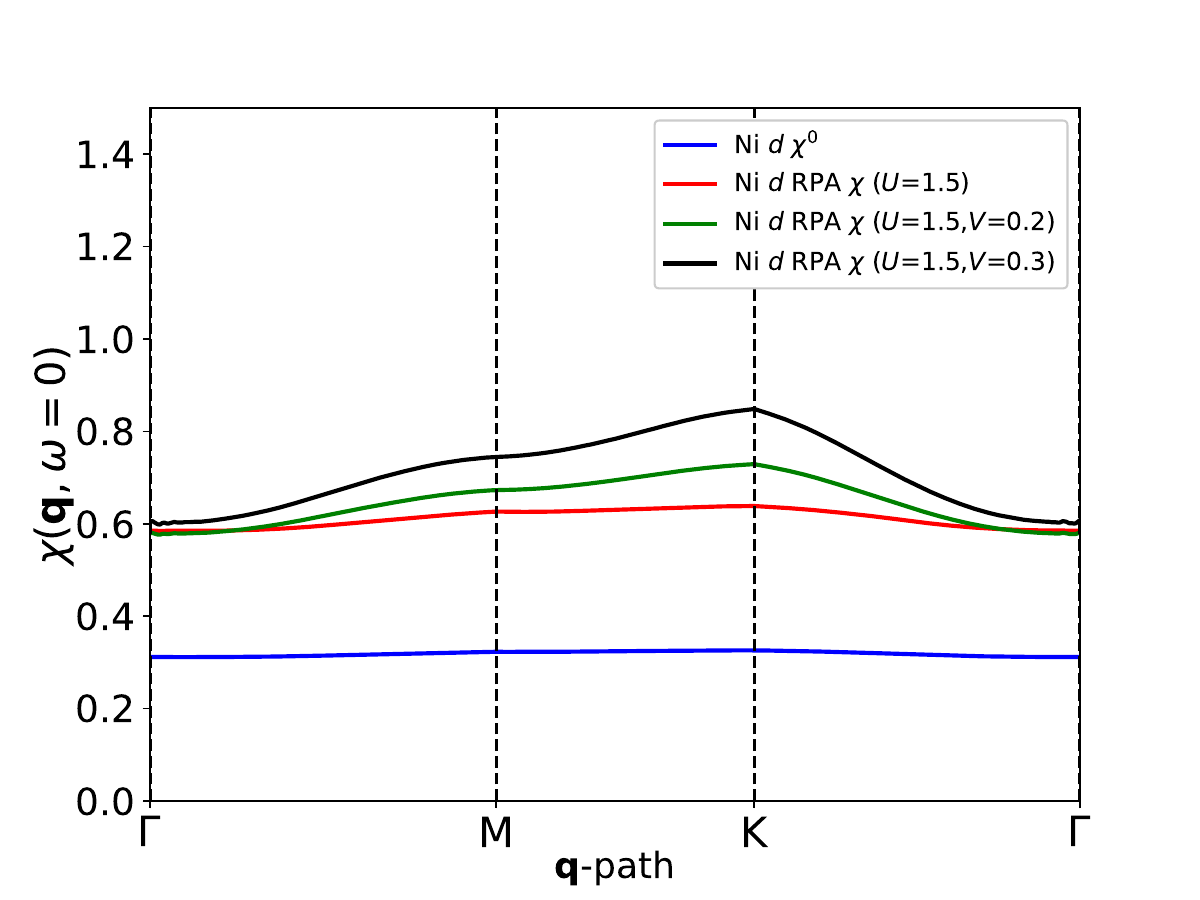}
\caption{The magnetic susceptibility for Ni$_{1/3}$NbS$_2$ at the doping $\delta$=2.0, the polarizability $\chi^0$ (left panel) and the RPA $\chi$ (right panel), computed using the form factor in Eq.\:\ref{eq:delta_form2}.
}
\label{fig:Ni_chi_no_matrix}
\end{figure*}

While the magnetic susceptibility in the main text is computed using the Form factor in Eq.\:\ref{eq:form_final}, the Nb $d_{z^2}$ $\chi^0$ profile can change notably once the matrix element effect of the Form factor is simplified using the delta function form in Eq.\:\ref{eq:delta_form2} for the susceptibility calculation. 
In Co$_{1/3}$NbS$_2$, the notable change occurs at the $M$ point where the peak height is slightly reduced and becomes almost degenerate with the peak at the $K$ point (see Fig.\:\ref{fig:Co_chi_no_matrix}).
This shows that the matrix element effect in Eq.\:\ref{eq:form_final} can favor the nesting along the $M$ direction as the Nb $d_{z^2}$ orbital modulates along the $M$ direction without the hybridization with Co $d$ orbitals.
However, the localized Co $d$ orbitals have no changes due to the matrix element effect as their profile is momentum independent and not affected by the hybridization effect.
In both Fe$_{1/3}$NbS$_2$ and Ni$_{1/3}$NbS$_2$ cases, the Nb $d_{z^2}$ susceptibility peak favors the $K$
 point in the BZ when the form factor is simplified using Eq.\:\ref{eq:delta_form2} (see Fig.\:\ref{fig:Fe_chi_no_matrix} and Fig.\:\ref{fig:Ni_chi_no_matrix}).
 
\end{document}